\documentclass[aps,prd,reprint,superscriptaddress]{revtex4-2}
\usepackage[english]{babel}
\usepackage{amsmath}
\usepackage{amssymb}
\usepackage{graphicx}
\usepackage{color}
\usepackage{lineno}
\newcommand{\SAP}{Dipartimento di Fisica - Sapienza Universit\`{a} di Roma, Piazzale Aldo Moro 2, 00185, Roma - Italy}
\newcommand{\SAPINF}{Dipartimento di Ingegneria Meccanica e Aerospaziale - Sapienza Universit\`{a} di Roma, Via Eudossiana 18, 00184, Roma - Italy}
\newcommand{\INFNRM}{INFN - Sezione di Roma, Piazzale Aldo Moro 2, 00185, Roma - Italy}
\newcommand{\INFNFE}{INFN - Sezione di Ferrara, via Saragat,1 44121, Ferrara - Italy}
\newcommand{\UNIFE}{Dipartimento di Fisica e Scienze della Terra, Università di Ferrara, Via Saragat 1, 44100, Ferrara, Italy}
\newcommand{\UNIFEN}{Dipartimento di Neuroscienze e Riabilitazione, Universit\`{a} di Ferrara, Via Luigi Borsari 46, 44121 Ferrara, Italy}
\newcommand{\NEEL}{Univ. Grenoble Alpes, CNRS, Grenoble INP, Institut N\'eel, 38000 Grenoble, France}

\begin{document}
 \title{Low-energy spectrum of the BULLKID detector array operated on surface}

\author{D.~Delicato}\email{daniele.delicato@roma1.infn.it}\affiliation{\NEEL}\affiliation{\SAP}\affiliation{\INFNRM}
\author{A.~Ahmad}\affiliation{\SAPINF}\affiliation{\INFNRM}
\author{L.~Bandiera}\affiliation{\INFNFE}
\author{M.~Calvo}\affiliation{\NEEL}
\author{M.~Cappelli}\affiliation{\SAP}\affiliation{\INFNRM}
\author{G.~Del~Castello}\affiliation{\SAP}\affiliation{\INFNRM}
\author{M.~del~Gallo~Roccagiovine}\affiliation{\SAP}\affiliation{\INFNRM}
\author{M.~Giammei}\affiliation{\SAP}\affiliation{\INFNRM}
\author{V.~Guidi}\affiliation{\UNIFE}\affiliation{\INFNFE}
\author{D.~Maiello}\affiliation{\SAP}\affiliation{\INFNRM}
\author{V.~Pettinacci}\affiliation{\INFNRM}
\author{M.~Romagnoni}\affiliation{\UNIFE}\affiliation{\INFNFE}
\author{M.~Tamisari}\affiliation{\UNIFEN}\affiliation{\INFNFE}
\author{A.~Cruciani}\affiliation{\INFNRM}
\author{A.~Mazzolari}\affiliation{\UNIFE}\affiliation{\INFNFE}
\author{A.~Monfardini}\affiliation{\NEEL}
\author{M.~Vignati}\email{marco.vignati@roma1.infn.it}\affiliation{\SAP}\affiliation{\INFNRM}

\begin{abstract}
We present the first continuous operation in a surface lab of BULLKID, a detector for searches of light Dark Matter and precision measurements of the coherent and elastic neutrino-nucleus scattering. 
The detector consists of an array of 60 cubic silicon particle absorbers of 0.34~g each, sensed by cryogenic kinetic inductance detectors.
The data presented focusses on one of the central elements of the array and on its surrounding elements used as veto. 
The energy spectrum resulting from an exposure of 39~hours to ambient backgrounds, obtained without radiation shields, is flat at the level of $(2.0\pm0.1\,{\rm stat.}\pm0.2\,{\rm syst.})\times10^6$~counts~/~keV~kg~days down to the energy threshold of $160\pm13$~eV.
The data analysis demonstrates the unique capability of rejecting backgrounds generated from interactions in other sites of the array, stemming from the segmented and monolithic structure of the detector. 
\end{abstract}

\maketitle

\section{INTRODUCTION}

Particle-physics experiments searching for weak or rare signals, such as the interaction of dark matter particles  with ordinary matter~\cite{Bertone2005} or the neutrino coherent and elastic scattering off atomic nuclei (CE$\nu$NS)~\cite{Freedman:1973yd,Akimov:2017ade}, need to be sensitive to small energy deposits with high target masses and negligible backgrounds. 
Experiments using cryogenic particle detectors, thanks to their high sensitivity to small energy depositions, are particularly suited to measure the sub-keV nuclear recoils following the scattering of dark matter particles with light mass~\cite{Bertone2018}, below 1~GeV/c$^2$~\cite{SuperCDMS:2017mbc,CRESST2019,EDELWEISS:2016boq}, or the scattering of low-energy neutrinos produced by nuclear reactors~\cite{Strauss:2017cuu,MINER}.

One of the challenges in cryogenic experiments consists in deploying large targets. Masses from fractions of grams~\cite{StraussGram,CRESST:2022lqw} or grams~\cite{CRAB2023} to tens of grams~\cite{RICOCHETAG,SuperCDMSCPD,CRESST2019} have been deployed so far, while masses of 1~kg or larger would allow to explore a much more significant space of the parameters. In the dark matter case they would enable sensitivities down to the background from solar neutrinos~\cite{appec}, while in the CE$\nu$NS case they would allow percent precisions on the measurement of the cross-section, enabling searches for new physics~\cite{Dodd:1991ni, Barranco:2005yy, Formaggio:2011jt, Dutta:2015nlo, Lindner:2016wff}.
  
Controlling the background is also proving to be a challenge. Even if radiation shields are adopted to mitigate the effects of environmental radioactivity, non-understood excess events populate the energy spectrum close to threshold (for a comprehensive summary see Ref.~\cite{excess2022}).
Hypotheses on the origin of these events point to solid state effects producing phonons mimicking the signals of particle interactions, such as mechanical stresses induced by the detector holders and lattice relaxations after cool down, or phonon signals generated by high-energy particle interactions in the inert materials in contact with the detector.

BULLKID~\cite{bullkid2022} is a new type of particle detector composed of an array of cryogenic sensors, designed to be scalable to high masses and to feature low-background thanks to the absence of inert material in contact with the single elements and to the high segmentation. In this paper we present the first continuous operation in the Laboratory of Cryogenic Detectors at Sapienza University and the measurement of the ambient background from 160~eV up to 1.2 keV.
After describing the detector and the experimental apparatus, we will describe the data analysis procedure, which introduces new methods to efficiently reduce the background. Finally we will interpret the measured energy spectrum and envision the prospects for next experiments. 

\section{EXPERIMENTAL SETUP}

\begin{figure}[t]
\centering
\includegraphics[width=\linewidth]{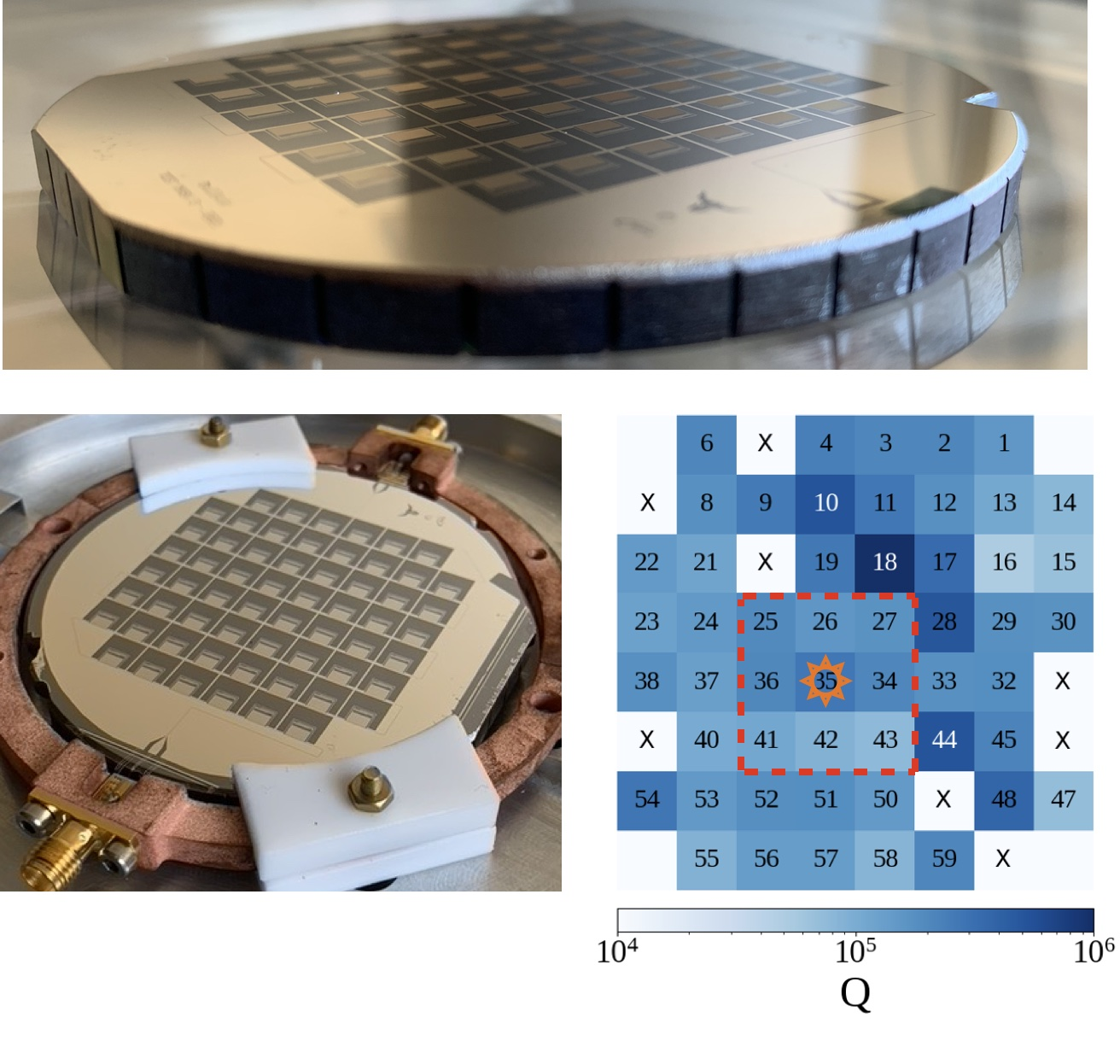}
\caption{
\textbf{Top:} BULLKID array: carved silicon wafer of 3" diameter and 5~mm thickness. A square grid of grooves creates 64 dice of 5.4x5.4x5~mm$^3$, and leaves intact a 0.5~mm thick common disk, which hosts the lithography of the KID phonon sensors;
\textbf{Bottom-left:} The array installed in the copper holder and held  by Teflon supports. The electrical feedline runs through all the KIDs for parallel readout;
\textbf{Bottom-right:} Quality factor $Q$ of each KID resonator in the array. The elements labelled as ``x'' are missing (see Appendix~\ref{sec:appendix1}). The  elements in the red-dashed contour are used in this work for the analysis of the energy spectrum, with the central die being the unit of interest (main) and the eight surrounding dice used as veto (sides).}
\label{fig1}
\end{figure}

BULLKID is an array of 60 silicon dice of 5.4x5.4x5 mm$^3$  and 0.34~g each, acting as particle absorbers and sensed by kinetic inductance detectors (KIDs), microwave resonators exploiting the kinetic inductance of superconductors~\cite{Day:2003fk}. The dice are carved out of a single  crystal, 3" in diameter and 5~mm thick, in order to obtain a monolithic structure. The carvings leave intact a 0.5~mm thick common disk that acts as both the holding structure of the dice and as substrate for the aluminum lithography of the KIDs (Fig.~\ref{fig1} top and bottom-left). 
KIDs are read in parallel, with a single feedline serving the entire array. 
The resulting device avoids the use of individual holding structures of the elements, which makes it a fully active particle detector. 
This, complemented with the high-segmentation, has been intended to provide additional background identification capabilities with respect to the state of the art, which consists in the use of independent array elements with individual and passive holders.

When particles interact in a silicon die their energy is converted to athermal phonons that scatter inside the crystalline lattice until part of them reach the KID interface where they are absorbed. The absorption of phonons induces a frequency shift of the resonator proportional to the energy released~\cite{swenson,moore1,cardani:2015tqa}. In order to sense the signal, the resonator is biased at the resonant frequency $f_0$ and the wave transmitted past the resonator is recorded with a heterodyne readout. Frequency shifts $\Delta f_0$ cause proportional variations to the phase of the transmitted wave, $\Delta\phi \simeq 4Q\, \Delta f_0/f_0$, where $Q$ is the quality factor of the resonator.

Figure~\ref{fig1} (bottom-right) shows the array map with the measured $Q$-factor of each KID, with a median value of $1.5\times10^5$ and a 68\% dispersion of $\pm0.7\times10^5$. 
In order to improve the uniformity of the response and to ease the combined analysis of different elements, the dispersion is reduced by a factor of 2 with respect to the first operation of the device  in Ref.~\cite{bullkid2022} (see Appendix~\ref{sec:appendix1} for technical details).

When phonons are generated in a die, part of them is absorbed by the KID coupled to it (main), while the rest leaks through the common disk in nearby dice and is absorbed by the respective KIDs (sides). With respect to the energy measured in the main die,  ($14\pm 3$)\% of energy was measured on average in each of the side KIDs in the vertical and horizontal directions and  ($5\pm1$)\% in those in the diagonal direction~\cite{bullkid2022}.
This effect is in principle unwanted, since it reduces the signal height in the main KID and thus its sensitivity. However it can be exploited to determine whether or not an event originates in the central die by comparing  the signal amplitude of the main KID with that of the side KIDs. 

For operation the array is installed in a copper holder by means of Teflon supports clamping the crystal in the peripheral region of the wafer, where there are no dice (Fig.~\ref{fig1} bottom-left). The holder is placed inside a shielding pot in aluminum, copper and Cryophy$^\text{\textregistered}$\cite{Cardani:2021wl}, providing protection from thermal radiation and external magnetic fields, and then is anchored to the coldest point of a dry $^3$He/$^4$He dilution refrigerator with base temperature of $20$~mK.

The readout is performed with an input coaxial line, running through the cryostat from the outside down to the device and attenuated at cryogenic temperature to reduce the noise temperature of the system, and with an output coaxial line from the device to the outside,  amplified by a HEMT low-noise amplifier thermalized at the 4 K stage of the cryostat. The microwaves to excite the KID resonators are generated and read back at room temperature by a commercial Ettus X300 board~\cite{ettus} operated with an open-source firmware~\cite{minutolo} customized for the needs of the experiment.

For the individual energy calibration,  optical fibers excited at room temperature by a 400~nm LED lamp shine controlled photon bursts to the die, on the face opposite to the KIDs~\cite{Cardani_2018}.
The systematic error associated to the energy calibration has been evaluated to be 8\% by comparing the results of independent scans performed at the beginning and at the end of the data taking.
The LED system is also used to provide a signal proxy to measure the phonon leakage across dice and to scan the detector response at lower energies for the evaluation of the trigger and analysis cuts efficiencies.

For this work we selected one of the 3$\times$3 clusters of KIDs with the most uniform $Q$ factor, KID 35 with its eight sides (see red contour in Fig.~\ref{fig1} bottom-right), which features a baseline resolution $\sigma_0=27\pm2$ eV. 

\begin{figure}[t]
\centering
\includegraphics[width=1.0\linewidth]{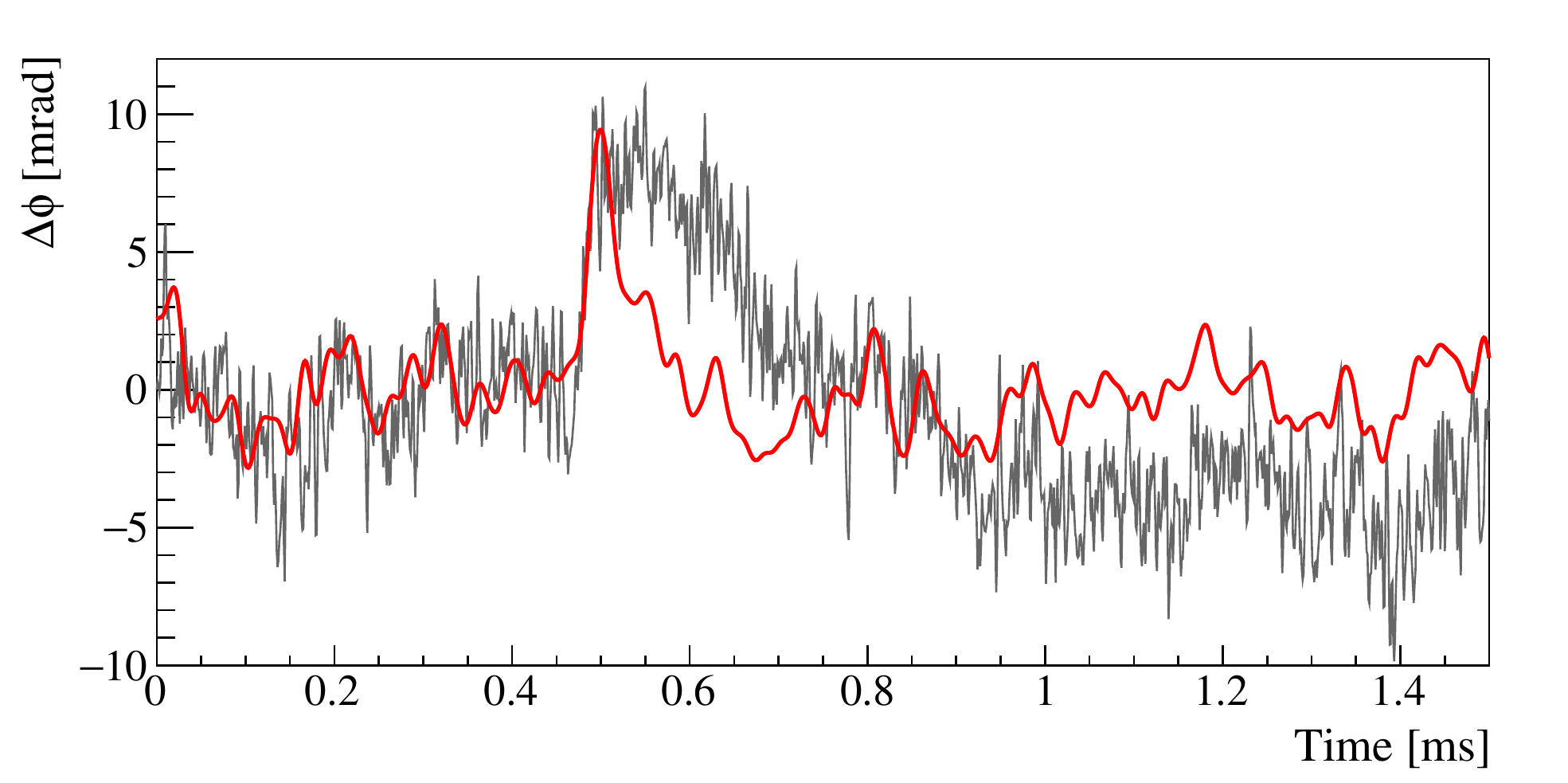}
\includegraphics[width=1.0\linewidth]{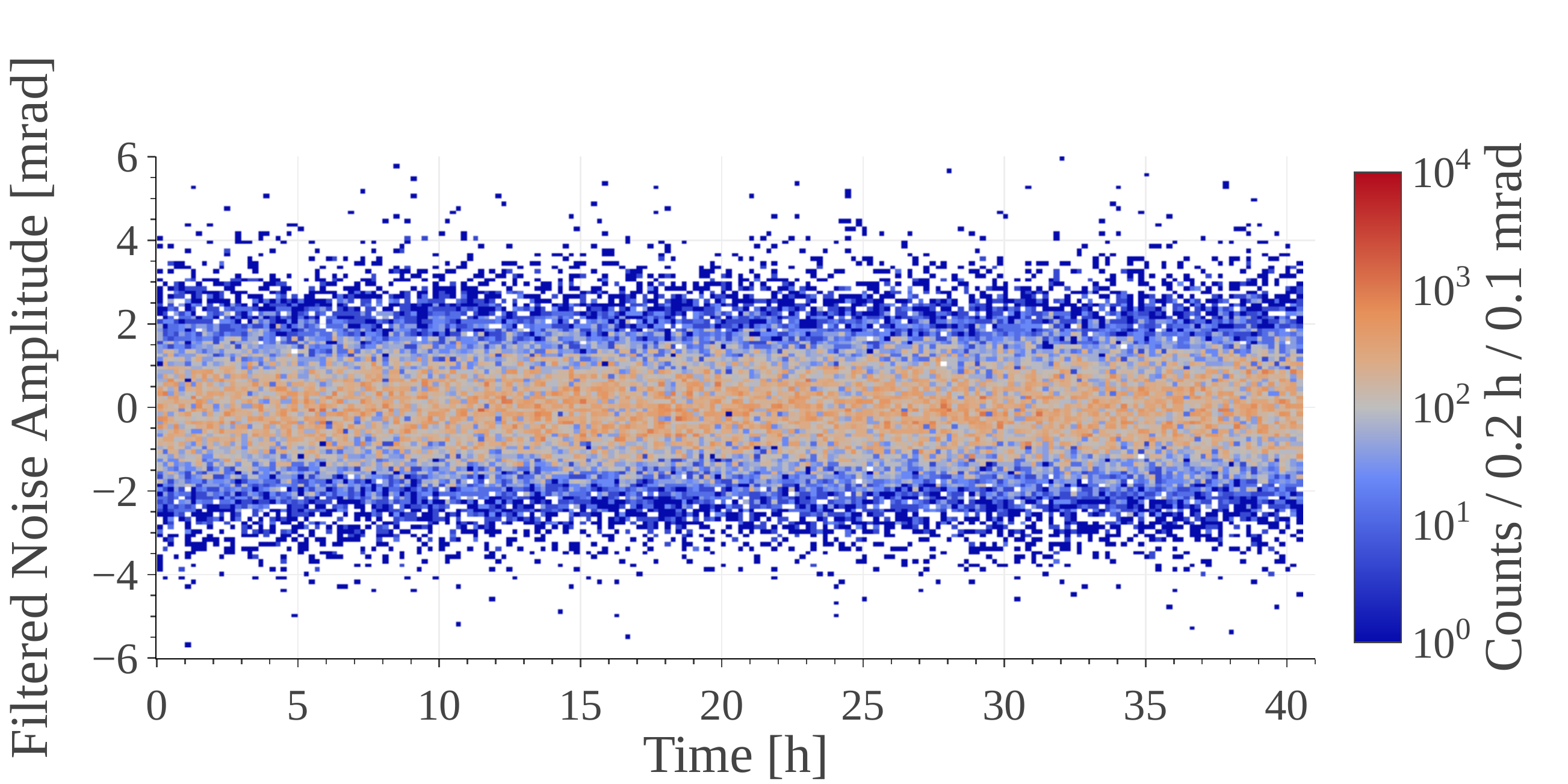}
\caption{{\bf Top:} Typical phase pulse recorded by the main KID (35) following a $\sim200$~eV energy release in the corresponding die (black) and the same waveform after application of the matched filter (red); {\bf Bottom:} Distribution of noise samples after matched filtering during the whole duration of the data taking. The RMS is 1.24 mrad and corresponds to 27 eV.} 
\label{fig2}
\end{figure}

During data taking the nine elements are acquired simultaneously, with the trigger running on the central element. The incoming data stream is filtered online using a matched filter~\cite{Radeka:1966,DiDomizio:2010ph}. Candidate signals are acquired when data exceed a threshold of $4\sigma_0$ (see a sample signal of $\sim200$~eV in Fig.~\ref{fig2} top). Noise samples are acquired at regular intervals in order to monitor the stability of the detector and did not show significant variation during the 40 hours of data taking (Fig.~\ref{fig2} bottom). Every hour the readout was stopped and restarted with 100~s dead time, yielding a net exposure of 39 hours.

At the end of the data taking we performed a measurement of 30 minutes with the trigger running on the negative side  of the data stream (reverse triggered data). This allows the study of the distribution of noise fluctuations mimicking real signals and to estimate the contribution of noise false positives to the energy spectrum close to threshold.

\section{DATA ANALYSIS AND RESULTS}

For the assessment of the energy spectrum we are interested in evaluating the signal amplitude, which is proportional to the energy released in the central die, in selecting events originating in the central die and in discarding those originating elsewhere.

\begin{figure}[t]
\centering
\includegraphics[width=1\linewidth]{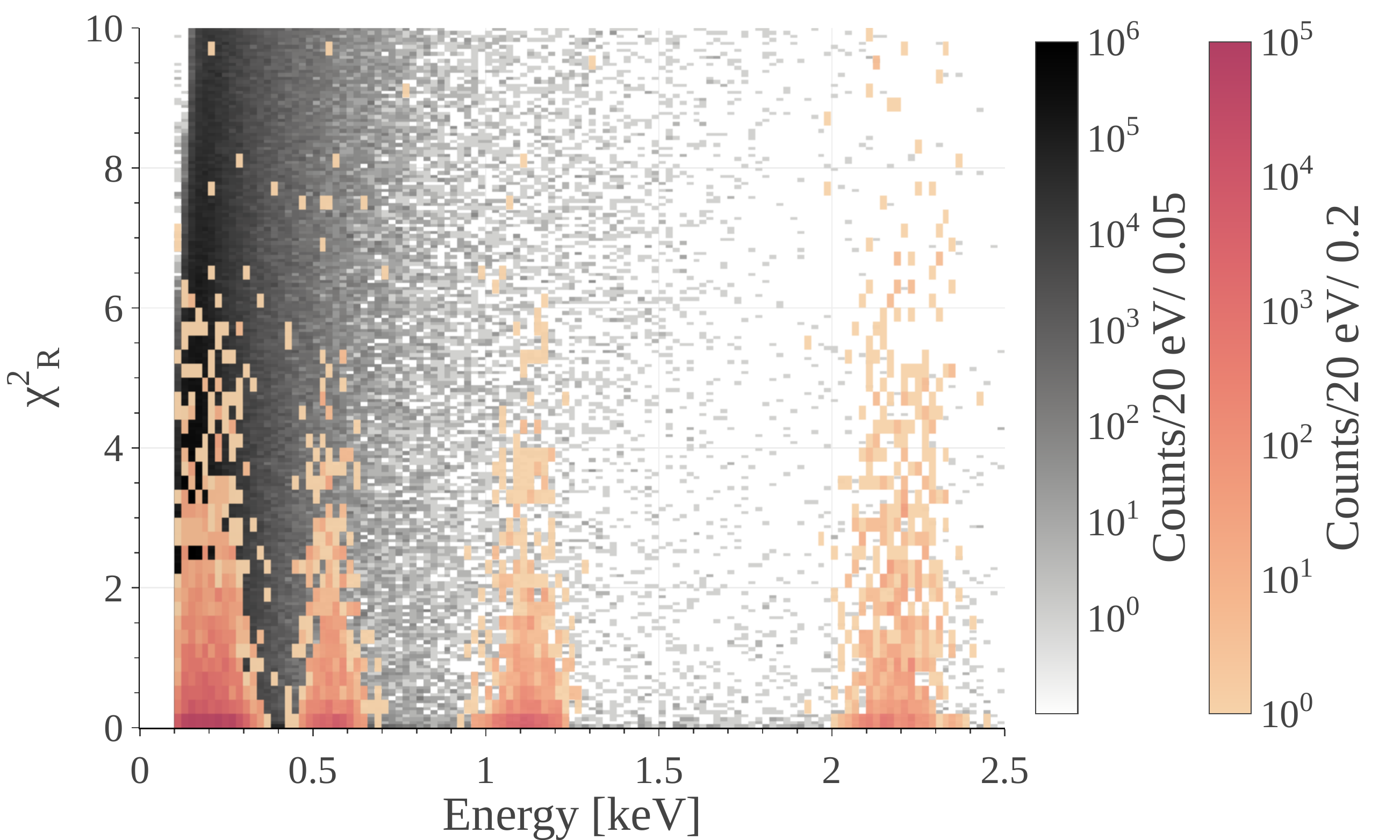}
\includegraphics[width=1\linewidth]{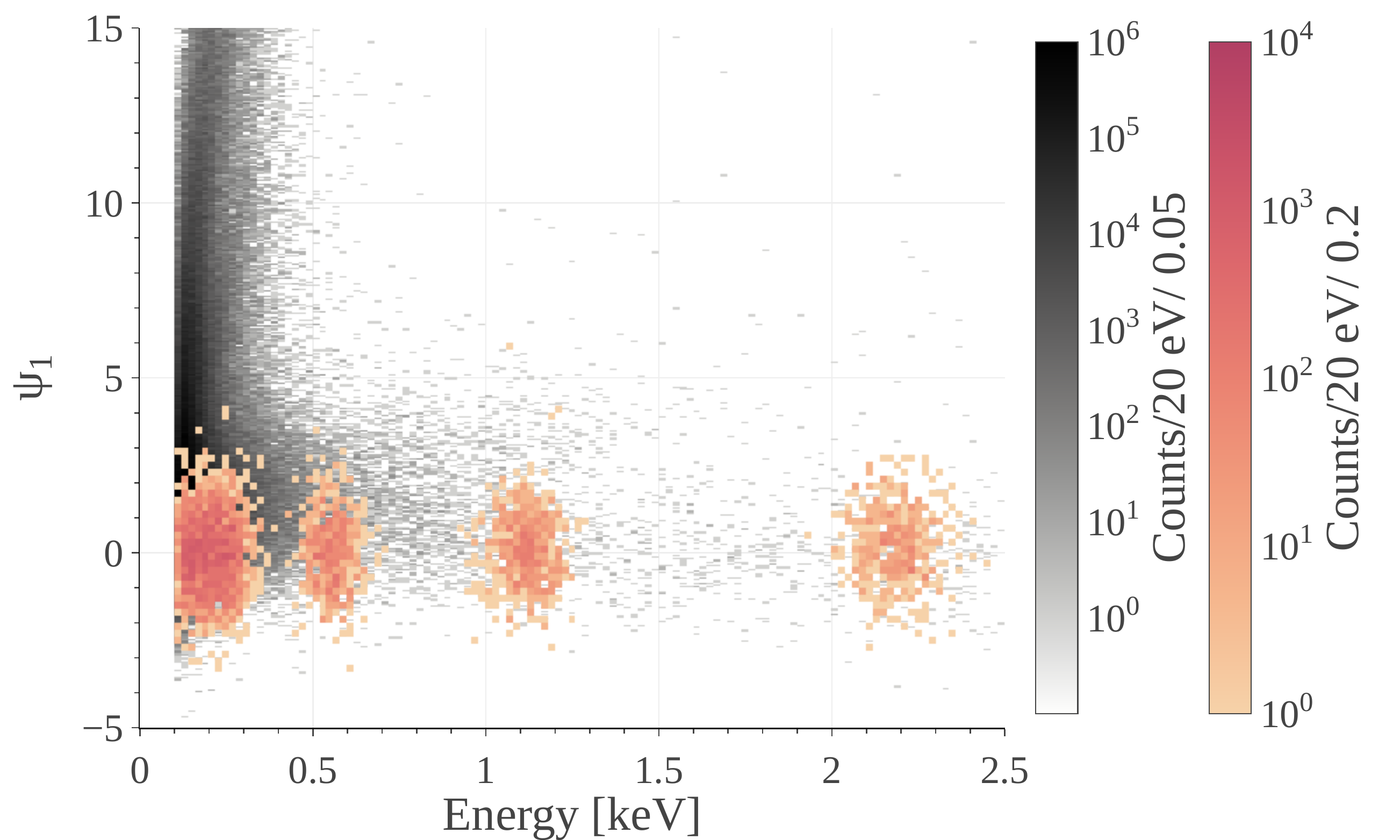}
\includegraphics[width=1\linewidth]{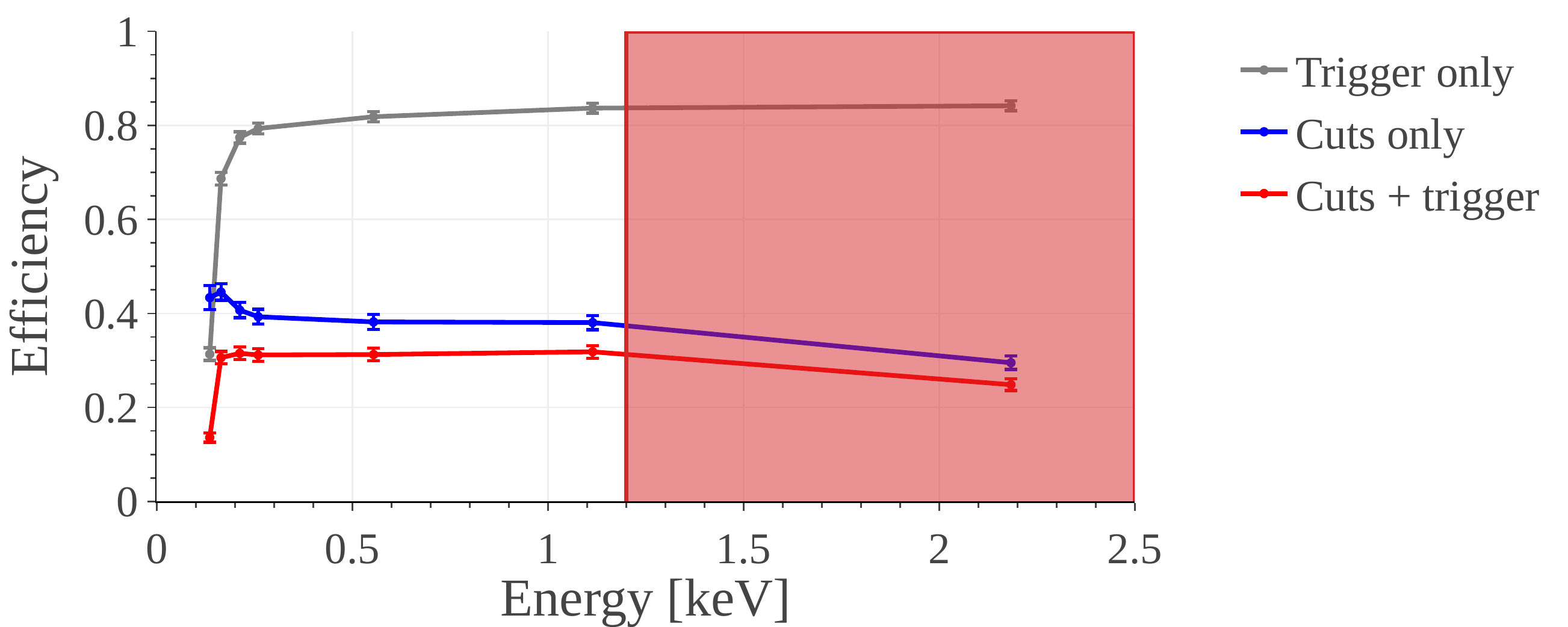}
\caption{{\bf Top:} $\chi^2_{R}$ as a function of energy in the main KID on all events (black) and on LED pulses (red colorscale). Events with $\chi^2_{L,R}<3$ are kept;
{\bf middle:} $\psi_{1}$, calculated from the amplitude of side 36 in Fig.~\protect\ref{fig1}, as a function of the energy in the main KID, after applying the cut on $\chi^2_{L,R}$. Events with $|\psi_{1}|<2$ are kept (same color coding of top panel);
{\bf bottom:} Trigger efficiency (gray), $\chi^2_{L,R}$ and $\psi_{1,...,8}$  cuts efficiency (blue) and total efficiency (red) computed on LED pulses. The region above 1.2 keV is excluded from the analysis (see discussion in the text).
}
\label{fig3}
\end{figure}

The triggered data are reprocessed offline with the matched filter and the signal amplitude is estimated as the maximum of the filtered waveform. The first event selection is made on the shape of the pulses on the main KID, by comparing it with the template signal shape estimated from LED pulses. We define the following parameter, which follows a $\chi^2$ distribution with 1 degree of freedom:
\begin{equation}\label{eq:chi}
\chi^2_i = \dfrac{\left[ S_i - a_i\cdot S_{i_{max}}\right]^2}{\sigma_0^2\cdot(1+a_i^2) -2a_i\cdot R(i-i_{max})}
\end{equation}
where $S_i$ is the filtered waveform at a generic sample $i$, $i_{max}$ the sample corresponding to the maximum of the waveform, $a_i = {S^{T}_i}/{S^{T}_{i_{max}}}$ is the unitary profile of the filtered template signal $S^T$, $R$ is the noise autocorrelation after filtering and $\sigma_{0}$ is the noise standard deviation of the main KID. We choose to evaluate $\chi^2_i$ at fixed  distance from $i_{max}$ in two points, corresponding to the half maximum of the filtered template on the left ($\chi^2_L$) and on the right ($\chi^2_R$) of the pulse maximum, respectively (see the distribution of the variable for all events and for LED events in Fig.~\ref{fig3} top). We choose to select events with $\chi^2_{L,R}<3$. From the combination of the two distributions, the selection is expected to keep $84$\,\% of the signal, not taking into account correlations or nonlinearities. 

The second level of event selection is the veto of events not originating in the central die. This is based on the analysis of the signal amplitude in side KIDs, which is evaluated in time coincidence with the maximum on the main KID~\footnote{No appreciable time delay is observed in the propagation of phonons between adjacent dice.}. For energy releases in the central die, the  ratio of the amplitude of the $n^{th}$ side pulse, $A_n$, relative to the amplitude of the main pulse, $A$, is estimated by shining pulses of LED light on the central die. By averaging hundreds of pulses we estimate this ratio as $r_n = \langle A_n / A \rangle$.
To disentangle events in the central die from those originating in other dice, we define the following variable:
\begin{equation}\label{eq:psi}
    \psi_n = \dfrac{A_n - A \cdot r_n}{\sqrt{\sigma_{0,n}^2 + r_n^2 \cdot \sigma_{0}^2}}
\end{equation}
where $\sigma_{0}$ ($\sigma_{0,n}$) is the noise standard deviation of the main ($n^{th}$) KID. 
This variable follows a standard normal distribution for events in the central die while its mean will be greater than zero for events in other dice (see the distribution of $\psi_1$, calculated from KID 36, for all events and for LED events in Fig.~\ref{fig3} middle). We choose to apply a cut $|\psi_{1,..,8}|<2.0$ which, from the combination of the 8 distributions, is expected to keep the $69$\% of the signal, not taking account  correlations or nonlinearities (sample signals on the nine KIDs before and after the application of cuts are reported in Appendix~\ref{sec:appendix2} for the interested reader).

The trigger and cut efficiencies are evaluated on the LED scan at energies of 140, 160, 210, 260, 550, 1100 and 2200 eV (Fig.~\ref{fig3} bottom). The trigger shows a plateau efficiency of 83\%, with a sharp cutoff while approaching the selected $4\sigma_0$ threshold of 110 eV. The efficiency of analysis cuts slightly decreases from threshold to higher energies because of non-linearities of the response of the KIDs. The combination of trigger and analysis efficiencies result in a constant efficiency of $( 32\pm2)$~\% from threshold up to $\sim$1.2 keV. We restrict the analysis to this region of the energy spectrum. Higher energies in any case would not be of interest for searches of light Dark Matter and of CE$\nu$NS at nuclear reactors. 

\begin{figure}[t]
\centering
\includegraphics[width=\linewidth]{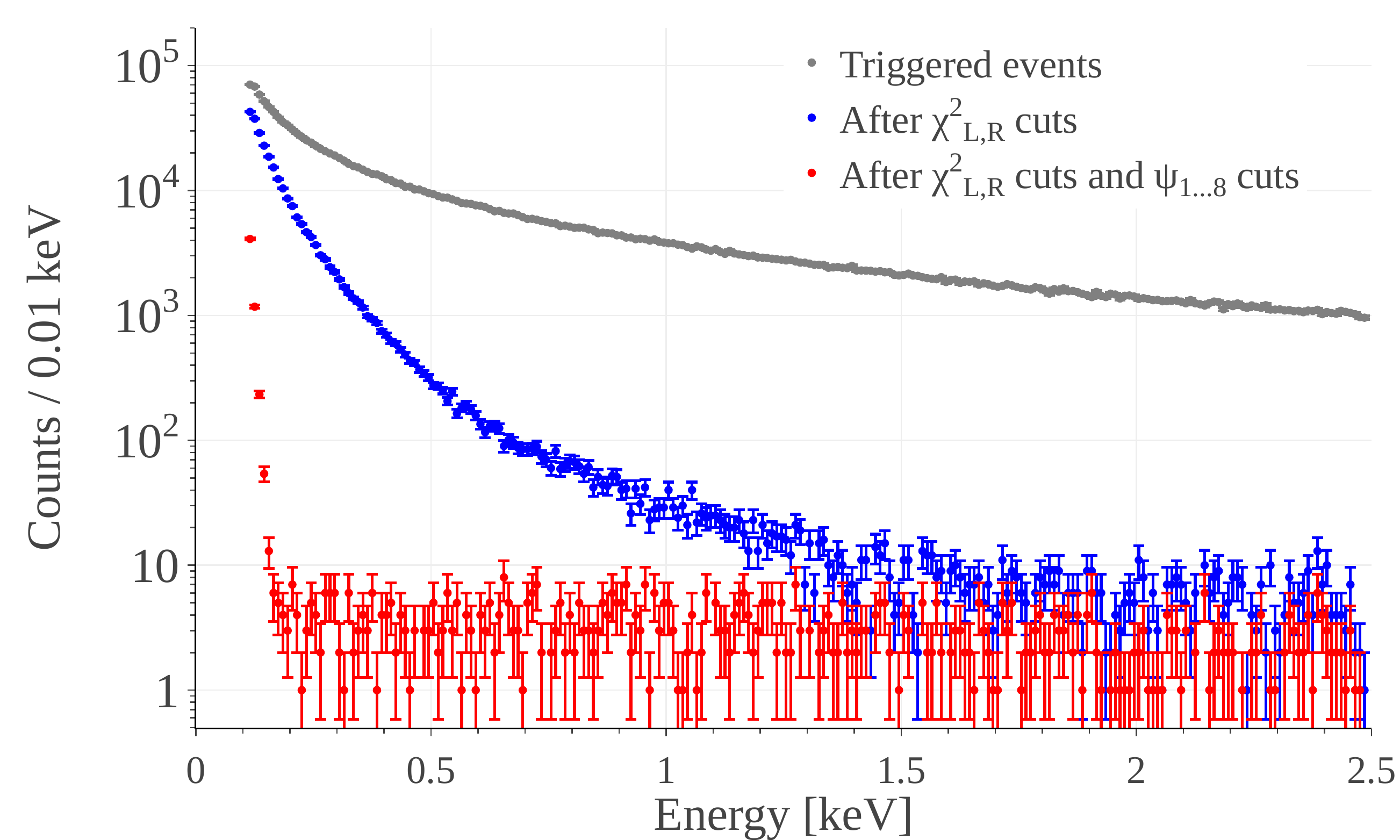}
\includegraphics[width=\linewidth]{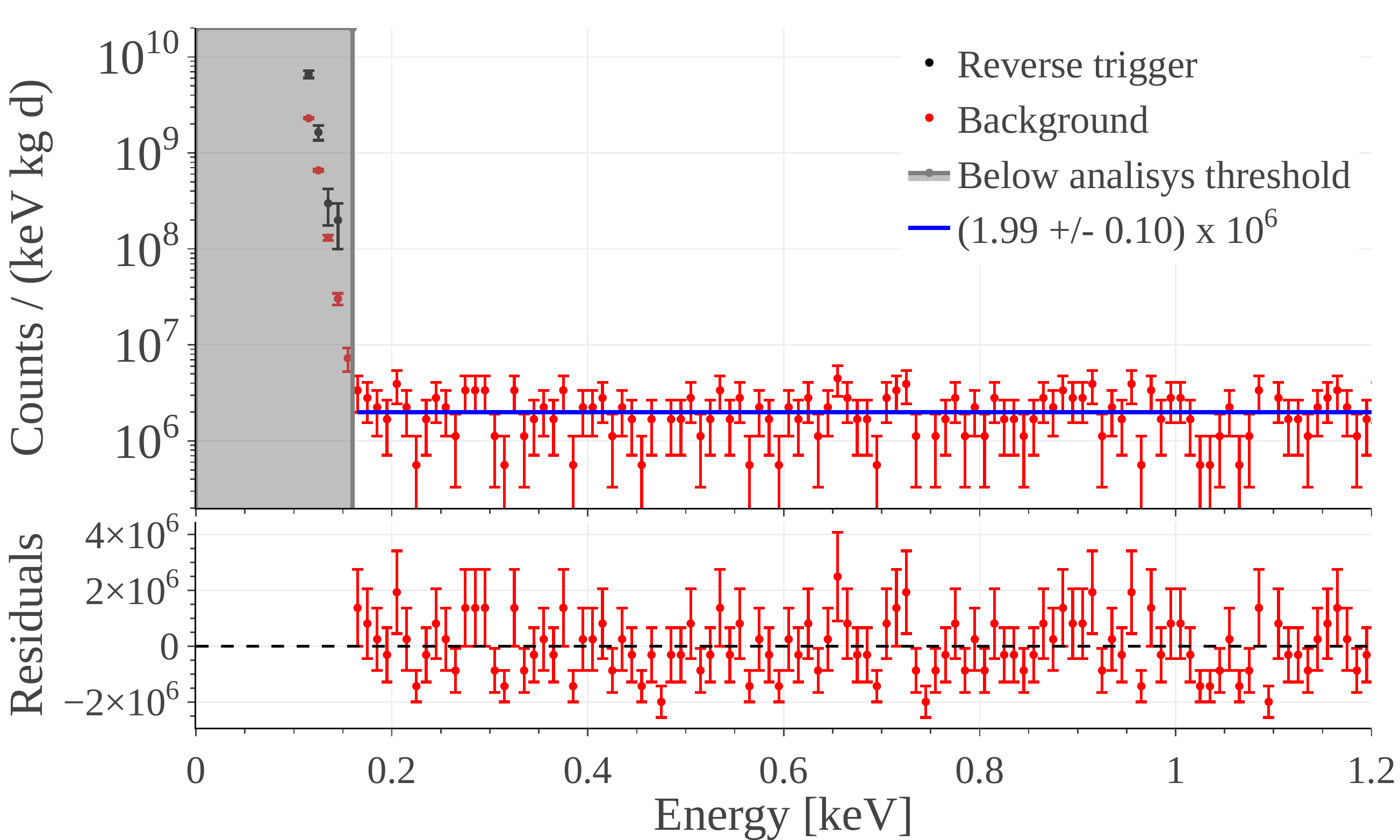}
\caption{ 
	{\bf Top:} Energy spectrum after 39~h of exposure of a single 0.34~g die to ambient backgrounds before cuts (gray), after application of cuts on the pulse shape variables $\chi^2_L$ and $\chi^2_R$ (blue) and after the additional application of cuts on all the coincidence variables $\psi_{1,..,8}$ (red). 
{\bf Bottom:} Energy spectral density with all cuts applied (red dots) and superimposed spectrum of the reverse triggered data (black dots).  
The rise below 160 eV follows that estimated from reverse triggered data (noise false-positive spectrum) and is excluded from the final analysis. 
The spectrum above 160~eV is flat with a density of $2.0\pm0.1\times10^ 6$~counts~/~keV~kg~day.
}
\label{fig4}
\end{figure}

Figure~\ref{fig4} (top) shows the energy spectrum without cuts, after the application of the pulse shape cuts and after the veto.  
The pulse shape cut reduces the event rate by more than two orders of magnitude in the middle of the spectrum, but is less effective at low energies. The additional application of the veto cut further reduces the rate while approaching the threshold.

Figure~\ref{fig4} (bottom) shows the resulting spectral density after including exposure and efficiencies.
The high counting rate above the trigger threshold is interpreted as due to noise false positives, as indicated by the energy spectrum of the reverse triggered data, which was derived following the same analysis procedure of standard data (see Appendix~\ref{sec:appendix3} for details). 
Since the reverse triggered data were acquired one day after the end of the data acquisition, the noise $\sigma_0$ was not exactly the same, instead being consistently 10\% larger. This caused an increase of the rate with respect to the normal data stream acquired before.

By restricting the analysis from an analysis threshold of 160~eV (corresponding to $6\sigma_0$) to 1.2~keV,
the spectrum is flat at a level of $(2.0\pm0.1)\times10^6$~counts~/~keV~kg~days (see the fit with a constant and the residuals in Fig.~\ref{fig4} bottom). The uncertainty on the energy scale end on the cut and trigger efficiencies contribute a systematic error amounting to $0.2\times10^6$~counts~/~keV~kg~days.

One may observe that the counting rate before cuts is several orders of magnitude higher than after cuts, which appears unnatural. This is a consequence of the central die being sensitive also to the leakage signal from interactions in the rest of the wafer, increasing the counting rate. Interactions happening anywhere in the wafer are seen as `quenched' by a KID sensor and populate the lower energies of its spectrum  increasing the spectral density. This ``background amplification'' at lower energies however does not represent a problem as long as leaking phonons can be efficiently tagged as shown in this work.

\section{CONCLUSION and PERSPECTIVES}
We presented 39 hours of continuous operation  in a surface lab of the BULLKID array of cryogenic detectors and we measured the ambient background at energies below 1.2~keV. Results show that it is possible to veto interactions not happening in the array element under examination down to the energy threshold of $160\pm13$~eV.  The presence of these events, which without vetoing would pollute the energy spectrum, is due to phonon cross-talk.
In our device this effect is favored by the presence of the common disk connecting the array elements but, thanks to the fully active structure, it can be identified and rejected. The spectrum obtained is flat, without the presence of events in excess unlike other experiments which already show a deviation from flatness at these energies (see Ref.~\cite{excess2022} and Appendix~\ref{sec:appendix4}). In this work the analysis was limited to an element of the array and to its 8 surrounding elements used as veto, but the presented technique could be even more enforced by the readout of the entire array of 60 elements.

This work represents only a first step towards background reduction. Given the operation on surface and given the absence of shields against ambient radioactivity, it is not possible to state whether the excess events seen by other experiments operated in lower background environments will still not be present.  

Next steps include to repeat this experiment in a shielded environment, yet to be identified, and to scale-up on the mass in view of a new GeV/sub-GeV/$c^2$ Dark Matter or CE$\nu$NS experiment. Our plan is to produce tens of BULLKID detectors, equal or larger than the one presented in this work, and to stack them to reach a total mass exceeding 0.5~kg. One of the advantages of this configuration would be a fully active volume of the stack, allowing for further background rejection by applying the same ``fiducialization’’ techniques as Dark Matter experiments using liquid scintillators~\cite{Xenon1TAnalysis,DarkSideCalibration} or single-site event discrimination techniques as in Double-$\beta$ decay experiments~\cite{CUOREAnalysis, GerdaMultiSite}. 
\vspace{0.5cm}
\\

\begin{acknowledgments}
This work was  supported by the INFN through the CSN5 grants BULLKID and BULLKID2 and by Sapienza University through the grant DANAE-TD. We acknowledge the support  of the PTA and Nanofab platforms for the fabrication of the devices and of the project HAMMER for the 3D-printed copper holder of the detector.
We thank L. Minutolo for the support on the electronics software.
We thank A. Girardi and M. Iannone of the INFN Sezione di Roma and V. Perino of Sapienza University for technical support.
\end{acknowledgments}

\bibliographystyle{apsrev4-2}
\bibliography{calder}

\begin{thebibliography}{39}%
\makeatletter
\providecommand \@ifxundefined [1]{%
 \@ifx{#1\undefined}
}%
\providecommand \@ifnum [1]{%
 \ifnum #1\expandafter \@firstoftwo
 \else \expandafter \@secondoftwo
 \fi
}%
\providecommand \@ifx [1]{%
 \ifx #1\expandafter \@firstoftwo
 \else \expandafter \@secondoftwo
 \fi
}%
\providecommand \natexlab [1]{#1}%
\providecommand \enquote  [1]{``#1''}%
\providecommand \bibnamefont  [1]{#1}%
\providecommand \bibfnamefont [1]{#1}%
\providecommand \citenamefont [1]{#1}%
\providecommand \href@noop [0]{\@secondoftwo}%
\providecommand \href [0]{\begingroup \@sanitize@url \@href}%
\providecommand \@href[1]{\@@startlink{#1}\@@href}%
\providecommand \@@href[1]{\endgroup#1\@@endlink}%
\providecommand \@sanitize@url [0]{\catcode `\\12\catcode `\$12\catcode
  `\&12\catcode `\#12\catcode `\^12\catcode `\_12\catcode `\%12\relax}%
\providecommand \@@startlink[1]{}%
\providecommand \@@endlink[0]{}%
\providecommand \url  [0]{\begingroup\@sanitize@url \@url }%
\providecommand \@url [1]{\endgroup\@href {#1}{\urlprefix }}%
\providecommand \urlprefix  [0]{URL }%
\providecommand \Eprint [0]{\href }%
\providecommand \doibase [0]{https://doi.org/}%
\providecommand \selectlanguage [0]{\@gobble}%
\providecommand \bibinfo  [0]{\@secondoftwo}%
\providecommand \bibfield  [0]{\@secondoftwo}%
\providecommand \translation [1]{[#1]}%
\providecommand \BibitemOpen [0]{}%
\providecommand \bibitemStop [0]{}%
\providecommand \bibitemNoStop [0]{.\EOS\space}%
\providecommand \EOS [0]{\spacefactor3000\relax}%
\providecommand \BibitemShut  [1]{\csname bibitem#1\endcsname}%
\let\auto@bib@innerbib\@empty
\bibitem [{\citenamefont {Bertone}\ \emph {et~al.}(2005)\citenamefont
  {Bertone}, \citenamefont {Hooper},\ and\ \citenamefont {Silk}}]{Bertone2005}%
  \BibitemOpen
  \bibfield  {author} {\bibinfo {author} {\bibfnamefont {G.}~\bibnamefont
  {Bertone}}, \bibinfo {author} {\bibfnamefont {D.}~\bibnamefont {Hooper}},\
  and\ \bibinfo {author} {\bibfnamefont {J.}~\bibnamefont {Silk}},\ }\bibfield
  {journal} {\bibinfo  {journal} {Phys. Rept.}\ }\textbf {\bibinfo {volume}
  {405}},\ \href {https://doi.org/https://doi.org/10.1017/CBO9780511770739}
  {https://doi.org/10.1017/CBO9780511770739} (\bibinfo {year}
  {2005})\BibitemShut {NoStop}%
\bibitem [{\citenamefont {Freedman}(1974)}]{Freedman:1973yd}%
  \BibitemOpen
  \bibfield  {author} {\bibinfo {author} {\bibfnamefont {D.~Z.}\ \bibnamefont
  {Freedman}},\ }\href {https://doi.org/10.1103/PhysRevD.9.1389} {\bibfield
  {journal} {\bibinfo  {journal} {Phys. Rev.}\ }\textbf {\bibinfo {volume}
  {D9}},\ \bibinfo {pages} {1389} (\bibinfo {year} {1974})}\BibitemShut
  {NoStop}%
\bibitem [{\citenamefont {Akimov}\ \emph {et~al.}(2017)\citenamefont {Akimov}
  \emph {et~al.}}]{Akimov:2017ade}%
  \BibitemOpen
  \bibfield  {author} {\bibinfo {author} {\bibfnamefont {D.}~\bibnamefont
  {Akimov}} \emph {et~al.} (\bibinfo {collaboration} {COHERENT}),\ }\href
  {https://doi.org/10.1126/science.aao0990} {\bibfield  {journal} {\bibinfo
  {journal} {Science}\ }\textbf {\bibinfo {volume} {357}},\ \bibinfo {pages}
  {1123} (\bibinfo {year} {2017})},\ \Eprint {https://arxiv.org/abs/1708.01294}
  {arXiv:1708.01294 [nucl-ex]} \BibitemShut {NoStop}%
\bibitem [{\citenamefont {Bertone}\ and\ \citenamefont
  {Tait}(2018)}]{Bertone2018}%
  \BibitemOpen
  \bibfield  {author} {\bibinfo {author} {\bibfnamefont {G.}~\bibnamefont
  {Bertone}}\ and\ \bibinfo {author} {\bibfnamefont {T.}~\bibnamefont {Tait}},\
  }\href {https://doi.org/https://doi.org/10.1038/s41586-018-0542-z} {\bibfield
   {journal} {\bibinfo  {journal} {Nature}\ }\textbf {\bibinfo {volume}
  {562}},\ \bibinfo {pages} {52} (\bibinfo {year} {2018})}\BibitemShut
  {NoStop}%
\bibitem [{\citenamefont {Agnese}\ \emph {et~al.}(2018)\citenamefont {Agnese},
  \citenamefont {Aramaki}, \citenamefont {Arnquist} \emph
  {et~al.}}]{SuperCDMS:2017mbc}%
  \BibitemOpen
  \bibfield  {author} {\bibinfo {author} {\bibfnamefont {R.}~\bibnamefont
  {Agnese}} \bibinfo {author} \emph
  {et~al.} (\bibinfo {collaboration} {SuperCDMS}),\ }\href
  {https://doi.org/10.1103/PhysRevLett.120.061802} {\bibfield  {journal}
  {\bibinfo  {journal} {Phys. Rev. Lett.}\ }\textbf {\bibinfo {volume} {120}},\
  \bibinfo {pages} {061802} (\bibinfo {year} {2018})}\BibitemShut {NoStop}%
\bibitem [{\citenamefont {Abdelhameed}\ \emph {et~al.}(2019)\citenamefont
  {Abdelhameed}, \citenamefont {Angloher}, \citenamefont {Bauer} \emph
  {et~al.}}]{CRESST2019}%
  \BibitemOpen
  \bibfield  {author} {\bibinfo {author} {\bibfnamefont {A.~H.}\ \bibnamefont
  {Abdelhameed}}
  \emph {et~al.},\ }\href@noop {} {\bibfield  {journal} {\bibinfo  {journal}
  {Phys. Rev. D}\ }\textbf {\bibinfo {volume} {100}},\ \bibinfo {pages}
  {102002} (\bibinfo {year} {2019})}\BibitemShut {NoStop}%
\bibitem [{\citenamefont {Armengaud}\ \emph {et~al.}(2016)\citenamefont
  {Armengaud}, \citenamefont {Arnaud}, \citenamefont {Augier} \emph
  {et~al.}}]{EDELWEISS:2016boq}%
  \BibitemOpen
  \bibfield  {author} {\bibinfo {author} {\bibfnamefont {E.}~\bibnamefont
  {Armengaud}} \emph {et~al.}
  (\bibinfo {collaboration} {EDELWEISS}),\ }\href
  {https://doi.org/10.1088/1475-7516/2016/05/019} {\bibfield  {journal}
  {\bibinfo  {journal} {JCAP}\ }\textbf {\bibinfo {volume} {05}},\ \bibinfo
  {pages} {019}}\BibitemShut {NoStop}%
\bibitem [{\citenamefont {Strauss}\ \emph
  {et~al.}(2017{\natexlab{a}})\citenamefont {Strauss}, \citenamefont {Rothe},
  \citenamefont {Angloher} \emph {et~al.}}]{Strauss:2017cuu}%
  \BibitemOpen
  \bibfield  {author} {\bibinfo {author} {\bibfnamefont {R.}~\bibnamefont
  {Strauss}} \emph
  {et~al.},\ }\href {https://doi.org/10.1140/epjc/s10052-017-5068-2} {\bibfield
   {journal} {\bibinfo  {journal} {Eur. Phys. J. C}\ }\textbf {\bibinfo
  {volume} {77}},\ \bibinfo {pages} {506} (\bibinfo {year}
  {2017}{\natexlab{a}})}\BibitemShut {NoStop}%
\bibitem [{\citenamefont {Agnolet}\ \emph {et~al.}(2017)\citenamefont
  {Agnolet}, \citenamefont {Baker}, \citenamefont {Barker} \emph
  {et~al.}}]{MINER}%
  \BibitemOpen
  \bibfield  {author} {\bibinfo {author} {\bibfnamefont {G.}~\bibnamefont
  {Agnolet}} \emph
  {et~al.},\ }\href
  {https://doi.org/https://doi.org/10.1016/j.nima.2017.02.024} {\bibfield
  {journal} {\bibinfo  {journal} {Nucl. Instr. Meth. in Phys. Res. A}\ }\textbf
  {\bibinfo {volume} {853}},\ \bibinfo {pages} {53} (\bibinfo {year}
  {2017})}\BibitemShut {NoStop}%
\bibitem [{\citenamefont {Strauss}\ \emph
  {et~al.}(2017{\natexlab{b}})\citenamefont {Strauss}, \citenamefont {Rothe},
  \citenamefont {Angloher} \emph {et~al.}}]{StraussGram}%
  \BibitemOpen
  \bibfield  {author} {\bibinfo {author} {\bibfnamefont {R.}~\bibnamefont
  {Strauss}} \emph
  {et~al.},\ }\href {https://doi.org/10.1103/PhysRevD.96.022009} {\bibfield
  {journal} {\bibinfo  {journal} {Phys. Rev. D}\ }\textbf {\bibinfo {volume}
  {96}},\ \bibinfo {pages} {022009} (\bibinfo {year}
  {2017}{\natexlab{b}})}\BibitemShut {NoStop}%
\bibitem [{\citenamefont {Angloher}\ \emph {et~al.}(2023)\citenamefont
  {Angloher} \emph {et~al.}}]{CRESST:2022lqw}%
  \BibitemOpen
  \bibfield  {author} {\bibinfo {author} {\bibfnamefont {G.}~\bibnamefont
  {Angloher}} \emph {et~al.} (\bibinfo {collaboration} {CRESST}),\ }\href
  {https://doi.org/10.1103/PhysRevD.107.122003} {\bibfield  {journal} {\bibinfo
   {journal} {Phys. Rev. D}\ }\textbf {\bibinfo {volume} {107}},\ \bibinfo
  {pages} {122003} (\bibinfo {year} {2023})},\ \Eprint
  {https://arxiv.org/abs/2212.12513} {arXiv:2212.12513 [astro-ph.CO]}
  \BibitemShut {NoStop}%
\bibitem [{\citenamefont {Abele}\ \emph {et~al.}(2023)\citenamefont {Abele}
  \emph {et~al.}}]{CRAB2023}%
  \BibitemOpen
  \bibfield  {author} {\bibinfo {author} {\bibfnamefont {H.}~\bibnamefont
  {Abele}} \emph {et~al.},\ }\href@noop {} {\bibfield  {journal} {\bibinfo
  {journal} {Phys. Rev. Lett.}\ }\textbf {\bibinfo {volume} {130}},\ \bibinfo
  {pages} {211802} (\bibinfo {year} {2023})}\BibitemShut {NoStop}%
\bibitem [{\citenamefont {Armengaud}\ \emph {et~al.}(2019)\citenamefont
  {Armengaud}, \citenamefont {Augier}, \citenamefont {Benoit} \emph
  {et~al.}}]{RICOCHETAG}%
  \BibitemOpen
  \bibfield  {author} {\bibinfo {author} {\bibfnamefont {E.}~\bibnamefont
  {Armengaud}} \emph {et~al.}
  (\bibinfo {collaboration} {EDELWEISS Collaboration}),\ }\href
  {https://doi.org/10.1103/PhysRevD.99.082003} {\bibfield  {journal} {\bibinfo
  {journal} {Phys. Rev. D}\ }\textbf {\bibinfo {volume} {99}},\ \bibinfo
  {pages} {082003} (\bibinfo {year} {2019})}\BibitemShut {NoStop}%
\bibitem [{\citenamefont {Alkhatib}\ \emph {et~al.}(2021)\citenamefont
  {Alkhatib}, \citenamefont {Amaral}, \citenamefont {Aralis} \emph
  {et~al.}}]{SuperCDMSCPD}%
  \BibitemOpen
  \bibfield  {author} {\bibinfo {author} {\bibfnamefont {I.}~\bibnamefont
  {Alkhatib}} \emph {et~al.}
  (\bibinfo {collaboration} {SuperCDMS Collaboration}),\ }\href
  {https://doi.org/10.1103/PhysRevLett.127.061801} {\bibfield  {journal}
  {\bibinfo  {journal} {Phys. Rev. Lett.}\ }\textbf {\bibinfo {volume} {127}},\
  \bibinfo {pages} {061801} (\bibinfo {year} {2021})}\BibitemShut {NoStop}%
\bibitem [{\citenamefont {Billard}\ and\ \citenamefont {others [APPEC
  Committee~Report]}(2022)}]{appec}%
  \BibitemOpen
  \bibfield  {author} {\bibinfo {author} {\bibfnamefont {J.}~\bibnamefont
  {Billard}}\ and\ \bibinfo {author} {\bibnamefont {others [APPEC
  Committee~Report]}},\ }\href@noop {} {\bibfield  {journal} {\bibinfo
  {journal} {Reports on Progress in Physics}\ }\textbf {\bibinfo {volume}
  {85}},\ \bibinfo {pages} {056201} (\bibinfo {year} {2022})}\BibitemShut
  {NoStop}%
\bibitem [{\citenamefont {Dodd}\ \emph {et~al.}(1991)\citenamefont {Dodd},
  \citenamefont {Papageorgiu},\ and\ \citenamefont {Ranfone}}]{Dodd:1991ni}%
  \BibitemOpen
  \bibfield  {author} {\bibinfo {author} {\bibfnamefont {A.~C.}\ \bibnamefont
  {Dodd}}, \bibinfo {author} {\bibfnamefont {E.}~\bibnamefont {Papageorgiu}},\
  and\ \bibinfo {author} {\bibfnamefont {S.}~\bibnamefont {Ranfone}},\ }\href
  {https://doi.org/10.1016/0370-2693(91)91064-3} {\bibfield  {journal}
  {\bibinfo  {journal} {Phys. Lett.}\ }\textbf {\bibinfo {volume} {B266}},\
  \bibinfo {pages} {434} (\bibinfo {year} {1991})}\BibitemShut {NoStop}%
\bibitem [{\citenamefont {Barranco}\ \emph {et~al.}(2005)\citenamefont
  {Barranco}, \citenamefont {Miranda},\ and\ \citenamefont
  {Rashba}}]{Barranco:2005yy}%
  \BibitemOpen
  \bibfield  {author} {\bibinfo {author} {\bibfnamefont {J.}~\bibnamefont
  {Barranco}}, \bibinfo {author} {\bibfnamefont {O.~G.}\ \bibnamefont
  {Miranda}},\ and\ \bibinfo {author} {\bibfnamefont {T.~I.}\ \bibnamefont
  {Rashba}},\ }\href {https://doi.org/10.1088/1126-6708/2005/12/021} {\bibfield
   {journal} {\bibinfo  {journal} {JHEP}\ }\textbf {\bibinfo {volume} {12}},\
  \bibinfo {pages} {021}},\ \Eprint {https://arxiv.org/abs/hep-ph/0508299}
  {arXiv:hep-ph/0508299 [hep-ph]} \BibitemShut {NoStop}%
\bibitem [{\citenamefont {Formaggio}\ \emph {et~al.}(2012)\citenamefont
  {Formaggio}, \citenamefont {Figueroa-Feliciano},\ and\ \citenamefont
  {Anderson}}]{Formaggio:2011jt}%
  \BibitemOpen
  \bibfield  {author} {\bibinfo {author} {\bibfnamefont {J.~A.}\ \bibnamefont
  {Formaggio}}, \bibinfo {author} {\bibfnamefont {E.}~\bibnamefont
  {Figueroa-Feliciano}},\ and\ \bibinfo {author} {\bibfnamefont {A.~J.}\
  \bibnamefont {Anderson}},\ }\href
  {https://doi.org/10.1103/PhysRevD.85.013009} {\bibfield  {journal} {\bibinfo
  {journal} {Phys. Rev.}\ }\textbf {\bibinfo {volume} {D85}},\ \bibinfo {pages}
  {013009} (\bibinfo {year} {2012})},\ \Eprint
  {https://arxiv.org/abs/1107.3512} {arXiv:1107.3512 [hep-ph]} \BibitemShut
  {NoStop}%
\bibitem [{\citenamefont {Dutta}\ \emph {et~al.}(2016)\citenamefont {Dutta},
  \citenamefont {Gao}, \citenamefont {Mahapatra}, \citenamefont {Mirabolfathi},
  \citenamefont {Strigari},\ and\ \citenamefont {Walker}}]{Dutta:2015nlo}%
  \BibitemOpen
  \bibfield  {author} {\bibinfo {author} {\bibfnamefont {B.}~\bibnamefont
  {Dutta}}  \emph {et~al.},\ }\href
  {https://doi.org/10.1103/PhysRevD.94.093002} {\bibfield  {journal} {\bibinfo
  {journal} {Phys. Rev.}\ }\textbf {\bibinfo {volume} {D94}},\ \bibinfo {pages}
  {093002} (\bibinfo {year} {2016})},\ \Eprint
  {https://arxiv.org/abs/1511.02834} {arXiv:1511.02834 [hep-ph]} \BibitemShut
  {NoStop}%
\bibitem [{\citenamefont {Lindner}\ \emph {et~al.}(2017)\citenamefont
  {Lindner}, \citenamefont {Rodejohann},\ and\ \citenamefont
  {Xu}}]{Lindner:2016wff}%
  \BibitemOpen
  \bibfield  {author} {\bibinfo {author} {\bibfnamefont {M.}~\bibnamefont
  {Lindner}}, \bibinfo {author} {\bibfnamefont {W.}~\bibnamefont
  {Rodejohann}},\ and\ \bibinfo {author} {\bibfnamefont {X.-J.}\ \bibnamefont
  {Xu}},\ }\href {https://doi.org/10.1007/JHEP03(2017)097} {\bibfield
  {journal} {\bibinfo  {journal} {JHEP}\ }\textbf {\bibinfo {volume} {03}},\
  \bibinfo {pages} {097}},\ \Eprint {https://arxiv.org/abs/1612.04150}
  {arXiv:1612.04150 [hep-ph]} \BibitemShut {NoStop}%
\bibitem [{\citenamefont {Adari}\ \emph {et~al.}(2022)\citenamefont {Adari}
  \emph {et~al.}}]{excess2022}%
  \BibitemOpen
  \bibfield  {author} {\bibinfo {author} {\bibfnamefont {P.}~\bibnamefont
  {Adari}} \emph {et~al.},\ }\href
  {https://doi.org/10.21468/SciPostPhysProc.9.001} {\bibfield  {journal}
  {\bibinfo  {journal} {SciPost Phys. Proc.}\ }\textbf {\bibinfo {volume}
  {009}},\ \bibinfo {pages} {001} (\bibinfo {year} {2022})}\BibitemShut
  {NoStop}%
\bibitem [{\citenamefont {Cruciani}\ \emph {et~al.}(2022)\citenamefont
  {Cruciani} \emph {et~al.}}]{bullkid2022}%
  \BibitemOpen
  \bibfield  {author} {\bibinfo {author} {\bibfnamefont {A.}~\bibnamefont
  {Cruciani}} \emph {et~al.},\ }\href {https://doi.org/10.1063/5.0128723}
  {\bibfield  {journal} {\bibinfo  {journal} {Appl. Phys. Lett.}\ }\textbf
  {\bibinfo {volume} {121}},\ \bibinfo {pages} {213504} (\bibinfo {year}
  {2022})}\BibitemShut {NoStop}%
\bibitem [{\citenamefont {Day}\ \emph {et~al.}(2003)\citenamefont {Day},
  \citenamefont {LeDuc}, \citenamefont {Mazin}, \citenamefont {Vayonakis},\
  and\ \citenamefont {Zmuidzinas}}]{Day:2003fk}%
  \BibitemOpen
  \bibfield  {author} {\bibinfo {author} {\bibfnamefont {P.~K.}\ \bibnamefont
  {Day}}, \bibinfo {author} {\bibfnamefont {H.~G.}\ \bibnamefont {LeDuc}},
  \bibinfo {author} {\bibfnamefont {B.~A.}\ \bibnamefont {Mazin}}, \bibinfo
  {author} {\bibfnamefont {A.}~\bibnamefont {Vayonakis}},\ and\ \bibinfo
  {author} {\bibfnamefont {J.}~\bibnamefont {Zmuidzinas}},\ }\href
  {https://doi.org/10.1038/nature02037} {\bibfield  {journal} {\bibinfo
  {journal} {Nature}\ }\textbf {\bibinfo {volume} {425}},\ \bibinfo {pages}
  {817} (\bibinfo {year} {2003})}\BibitemShut {NoStop}%
\bibitem [{\citenamefont {Swenson}\ \emph {et~al.}(2010)\citenamefont
  {Swenson}, \citenamefont {Cruciani}, \citenamefont {Benoit} \emph
  {et~al.}}]{swenson}%
  \BibitemOpen
  \bibfield  {author} {\bibinfo {author} {\bibfnamefont {L.~J.}\ \bibnamefont
  {Swenson}} \emph
  {et~al.},\ }\href {https://doi.org/10.1063/1.3459142} {\bibfield  {journal}
  {\bibinfo  {journal} {{A}ppl. {P}hys. {L}ett.}\ }\textbf {\bibinfo {volume}
  {96}},\ \bibinfo {pages} {263511} (\bibinfo {year} {2010})}\BibitemShut
  {NoStop}%
\bibitem [{\citenamefont {Moore}\ \emph {et~al.}(2012)\citenamefont {Moore}
  \emph {et~al.}}]{moore1}%
  \BibitemOpen
  \bibfield  {author} {\bibinfo {author} {\bibfnamefont {D.~C.}\ \bibnamefont
  {Moore}} \emph {et~al.},\ }\href {https://doi.org/10.1007/s10909-011-0434-1}
  {\bibfield  {journal} {\bibinfo  {journal} {J. Low Temp. Phys.}\ }\textbf
  {\bibinfo {volume} {167}},\ \bibinfo {pages} {329} (\bibinfo {year}
  {2012})}\BibitemShut {NoStop}%
\bibitem [{\citenamefont {Cardani}\ \emph {et~al.}(2015)\citenamefont
  {Cardani}, \citenamefont {Colantoni}, \citenamefont {Cruciani}, \citenamefont
  {Di~Domizio}, \citenamefont {Vignati}, \citenamefont {Bellini}, \citenamefont
  {Casali}, \citenamefont {Castellano}, \citenamefont {Coppolecchia},
  \citenamefont {Cosmelli} \emph {et~al.}}]{cardani:2015tqa}%
  \BibitemOpen
  \bibfield  {author} {\bibinfo {author} {\bibfnamefont {L.}~\bibnamefont
  {Cardani}} \emph {et~al.},\ }\href@noop {} {\bibfield
  {journal} {\bibinfo  {journal} {Appl.Phys.Lett.}\ }\textbf {\bibinfo {volume}
  {107}},\ \bibinfo {pages} {093508} (\bibinfo {year} {2015})},\ \Eprint
  {https://arxiv.org/abs/1505.04666} {arXiv:1505.04666} \BibitemShut {NoStop}%
\bibitem [{\citenamefont {Cardani}\ \emph {et~al.}(2021)\citenamefont
  {Cardani}, \citenamefont {Casali}, \citenamefont {Colantoni} \emph
  {et~al.}}]{Cardani:2021wl}%
  \BibitemOpen
  \bibfield  {author} {\bibinfo {author} {\bibfnamefont {L.}~\bibnamefont
  {Cardani}} \emph
  {et~al.},\ }\href {https://doi.org/10.1140/epjc/s10052-021-09454-5}
  {\bibfield  {journal} {\bibinfo  {journal} {Eur. Phys. J. C}\ }\textbf
  {\bibinfo {volume} {81}},\ \bibinfo {pages} {636} (\bibinfo {year}
  {2021})}\BibitemShut {NoStop}%
\bibitem [{ett()}]{ettus}%
  \BibitemOpen
  \href@noop {} {\bibinfo {title} {https://www.ettus.com}}\BibitemShut
  {NoStop}%
\bibitem [{\citenamefont {Minutolo}\ \emph {et~al.}(2019)\citenamefont
  {Minutolo}, \citenamefont {Steinbach}, \citenamefont {Wandui},\ and\
  \citenamefont {O'Brient}}]{minutolo}%
  \BibitemOpen
  \bibfield  {author} {\bibinfo {author} {\bibfnamefont {L.}~\bibnamefont
  {Minutolo}}, \bibinfo {author} {\bibfnamefont {B.}~\bibnamefont {Steinbach}},
  \bibinfo {author} {\bibfnamefont {A.}~\bibnamefont {Wandui}},\ and\ \bibinfo
  {author} {\bibfnamefont {R.}~\bibnamefont {O'Brient}},\ }\href
  {https://doi.org/10.1109/TASC.2019.2912027} {\bibfield  {journal} {\bibinfo
  {journal} {IEEE Trans. Appl. Supercond.}\ }\textbf {\bibinfo {volume} {29}},\
  \bibinfo {pages} {1} (\bibinfo {year} {2019})}\BibitemShut {NoStop}%
\bibitem [{\citenamefont {Cardani}\ \emph {et~al.}(2018)\citenamefont
  {Cardani}, \citenamefont {Casali}, \citenamefont {Cruciani} \emph
  {et~al.}}]{Cardani_2018}%
  \BibitemOpen
  \bibfield  {author} {\bibinfo {author} {\bibfnamefont {L.}~\bibnamefont
  {Cardani}} \emph
  {et~al.},\ }\href {https://doi.org/10.1088/1361-6668/aac1d4} {\bibfield
  {journal} {\bibinfo  {journal} {Supercond. Sci. Technol.}\ }\textbf {\bibinfo
  {volume} {31}},\ \bibinfo {pages} {075002} (\bibinfo {year}
  {2018})}\BibitemShut {NoStop}%
\bibitem [{\citenamefont {Radeka}\ and\ \citenamefont
  {Karlovac}(1967)}]{Radeka:1966}%
  \BibitemOpen
  \bibfield  {author} {\bibinfo {author} {\bibfnamefont {V.}~\bibnamefont
  {Radeka}}\ and\ \bibinfo {author} {\bibfnamefont {N.}~\bibnamefont
  {Karlovac}},\ }\href@noop {} {\bibfield  {journal} {\bibinfo  {journal}
  {Nucl. Instrum. Methods}\ }\textbf {\bibinfo {volume} {52}},\ \bibinfo
  {pages} {86} (\bibinfo {year} {1967})}\BibitemShut {NoStop}%
\bibitem [{\citenamefont {Di~Domizio}\ \emph {et~al.}(2011)\citenamefont
  {Di~Domizio}, \citenamefont {Orio},\ and\ \citenamefont
  {Vignati}}]{DiDomizio:2010ph}%
  \BibitemOpen
  \bibfield  {author} {\bibinfo {author} {\bibfnamefont {S.}~\bibnamefont
  {Di~Domizio}}, \bibinfo {author} {\bibfnamefont {F.}~\bibnamefont {Orio}},\
  and\ \bibinfo {author} {\bibfnamefont {M.}~\bibnamefont {Vignati}},\
  }\href@noop {} {\bibfield  {journal} {\bibinfo  {journal} {JINST}\ }\textbf
  {\bibinfo {volume} {6}},\ \bibinfo {pages} {P02007}},\ \Eprint
  {https://arxiv.org/abs/1012.1263} {arXiv:1012.1263 [astro-ph.IM]}
  \BibitemShut {NoStop}%
\bibitem [{Note1()}]{Note1}%
  \BibitemOpen
  \bibinfo {note} {No appreciable time delay is observed in the propagation of
  phonons between adiacent dice.}\BibitemShut {Stop}%
\bibitem [{\citenamefont {Aprile}\ \emph {et~al.}(2019)\citenamefont {Aprile}
  \emph {et~al.}}]{Xenon1TAnalysis}%
  \BibitemOpen
  \bibfield  {author} {\bibinfo {author} {\bibfnamefont {E.}~\bibnamefont
  {Aprile}} \emph {et~al.} (\bibinfo {collaboration} {XENON Collaboration}),\
  }\href {https://doi.org/10.1103/PhysRevD.100.052014} {\bibfield  {journal}
  {\bibinfo  {journal} {Phys. Rev. D}\ }\textbf {\bibinfo {volume} {100}},\
  \bibinfo {pages} {052014} (\bibinfo {year} {2019})}\BibitemShut {NoStop}%
\bibitem [{\citenamefont {Agnes}\ \emph {et~al.}(2021)\citenamefont {Agnes}
  \emph {et~al.}}]{DarkSideCalibration}%
  \BibitemOpen
  \bibfield  {author} {\bibinfo {author} {\bibfnamefont {P.}~\bibnamefont
  {Agnes}} \emph {et~al.} (\bibinfo {collaboration} {DarkSide Collaboration}),\
  }\href {https://doi.org/10.1103/PhysRevD.104.082005} {\bibfield  {journal}
  {\bibinfo  {journal} {Phys. Rev. D}\ }\textbf {\bibinfo {volume} {104}},\
  \bibinfo {pages} {082005} (\bibinfo {year} {2021})}\BibitemShut {NoStop}%
\bibitem [{\citenamefont {Alduino}\ \emph {et~al.}(2016)\citenamefont {Alduino}
  \emph {et~al.}}]{CUOREAnalysis}%
  \BibitemOpen
  \bibfield  {author} {\bibinfo {author} {\bibfnamefont {C.}~\bibnamefont
  {Alduino}} \emph {et~al.} (\bibinfo {collaboration} {CUORE Collaboration}),\
  }\href {https://doi.org/10.1103/PhysRevC.93.045503} {\bibfield  {journal}
  {\bibinfo  {journal} {Phys. Rev. C}\ }\textbf {\bibinfo {volume} {93}},\
  \bibinfo {pages} {045503} (\bibinfo {year} {2016})}\BibitemShut {NoStop}%
\bibitem [{\citenamefont {Agostini}\ \emph {et~al.}(2013)\citenamefont
  {Agostini} \emph {et~al.}}]{GerdaMultiSite}%
  \BibitemOpen
  \bibfield  {author} {\bibinfo {author} {\bibfnamefont {M.}~\bibnamefont
  {Agostini}} \emph {et~al.},\ }\href
  {https://doi.org/10.1140/epjc/s10052-013-2583-7} {\bibfield  {journal}
  {\bibinfo  {journal} {Eur. Phys. J C}\ }\textbf {\bibinfo {volume} {73}},\
  \bibinfo {pages} {2583} (\bibinfo {year} {2013})}\BibitemShut {NoStop}%
\bibitem [{\citenamefont {Zmuidzinas}(2012)}]{zmu_annrev2012}%
  \BibitemOpen
  \bibfield  {author} {\bibinfo {author} {\bibfnamefont {J.}~\bibnamefont
  {Zmuidzinas}},\ }\href
  {https://doi.org/10.1146/annurev-conmatphys-020911-125022} {\bibfield
  {journal} {\bibinfo  {journal} {Annu.Rev.Cond.Mat.Phys.}\ }\textbf {\bibinfo
  {volume} {3}},\ \bibinfo {pages} {169} (\bibinfo {year} {2012})}\BibitemShut
  {NoStop}%
\bibitem [{\citenamefont {Lee}\ \emph {et~al.}(1995)\citenamefont {Lee},
  \citenamefont {Liu},\ and\ \citenamefont {Itoh}}]{CPWSlot}%
  \BibitemOpen
  \bibfield  {author} {\bibinfo {author} {\bibfnamefont {C.-Y.}\ \bibnamefont
  {Lee}}, \bibinfo {author} {\bibfnamefont {Y.}~\bibnamefont {Liu}},\ and\
  \bibinfo {author} {\bibfnamefont {T.}~\bibnamefont {Itoh}},\ }\href
  {https://doi.org/10.1109/22.475632} {\bibfield  {journal} {\bibinfo
  {journal} {IEEE Trans. Microw. Theory Tech.}\ }\textbf {\bibinfo {volume}
  {43}},\ \bibinfo {pages} {2759} (\bibinfo {year} {1995})}\BibitemShut
  {NoStop}%
\bibitem [{\citenamefont {Alduino}\ \emph {et~al.}(2017)\citenamefont {Alduino}
  \emph {et~al.}}]{CUORELowEnergyAnalysis}%
  \BibitemOpen
  \bibfield  {author} {\bibinfo {author} {\bibfnamefont {C.}~\bibnamefont
  {Alduino}} \emph {et~al.},\ }\href
  {https://doi.org/10.1140/epjc/s10052-017-5433-1} {\bibfield  {journal}
  {\bibinfo  {journal} {Eur. Phys. J C}\ }\textbf {\bibinfo {volume} {77}},\
  \bibinfo {pages} {857} (\bibinfo {year} {2017})}\BibitemShut {NoStop}%
\end{thebibliography}%

\begin{appendix}

\section{UNIFORMITY OF QUALITY FACTORS}\label{sec:appendix1}
The quality factor $Q$ of a superconducting resonator  depends on the coupling ($Q_c$) and internal ($Q_i$) quality factors as $Q^{-1} = Q_c^{-1} + Q_i^{-1}$. In this work
$Q_c$ is set by design to $1.5\times10^{5}$, while $Q_i$ depends on the quality of the aluminum film and on internal dissipation~\cite{zmu_annrev2012}. Thanks to the high quality of the production, $Q_i$ was found to be in excess of 1 million, impyling that $Q\simeq Q_c$.

The feedline of the array is a coplanar wave guide (CPW), where the signal runs in a central wire faced to lateral ground planes.
However, the propagation in the CPW of a slot mode alters the coupling at different positions along the line. A way to correct for this issue consist in performing several bondings along the feedline between the two ground planes (also called air-bridges), in order to fix their potential and to prevent the occurrence of the slot mode~\cite{CPWSlot}. 

\begin{figure}[t]
\includegraphics[width=0.8\linewidth]{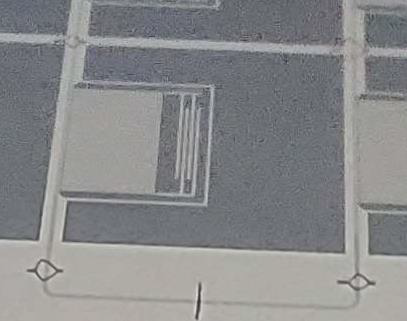}
\includegraphics[width=1.0\linewidth]{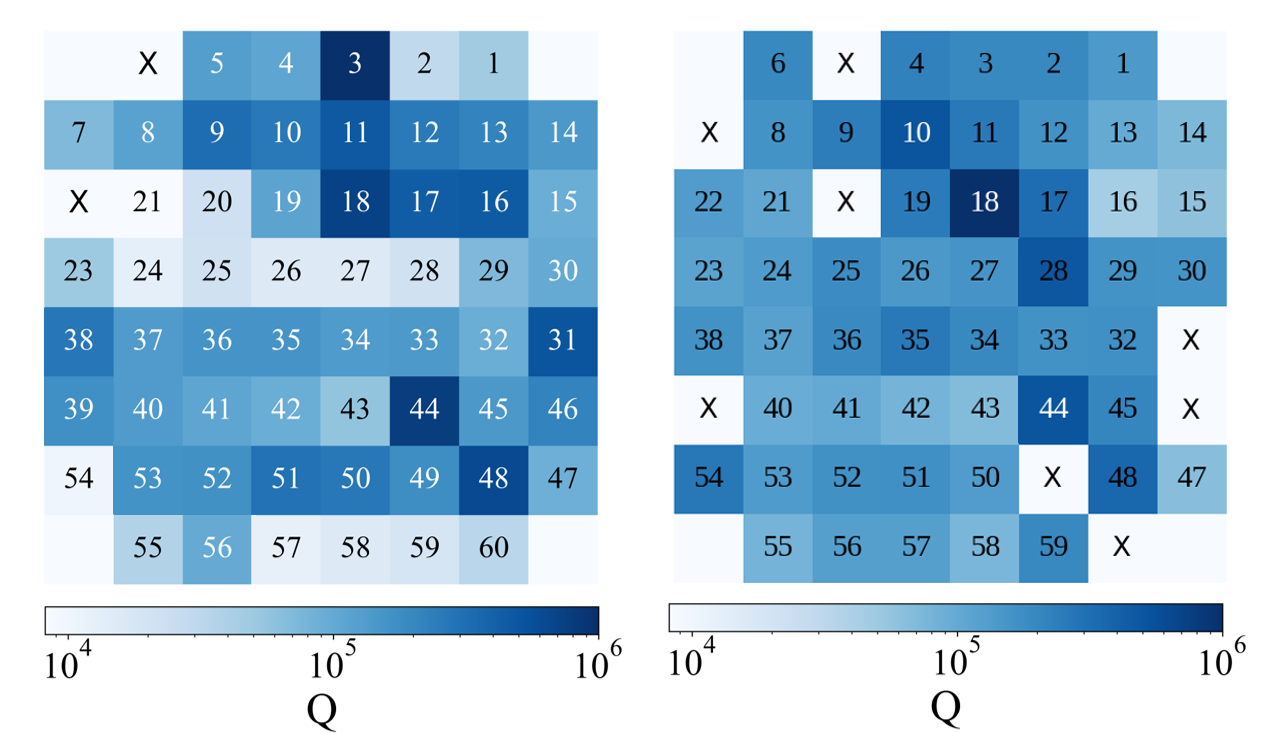}
\caption{{\bf Top:} Detail of the air-bridges connecting the two ground planes of the CPW in order to improve uniformity of the coupling quality factor $Q_c$. The air-bridges are repeated every 6~mm, in correspondence of each groove in the wafer.
{\bf bottom left:} Distribution of the $Q$ factor before applying the air bridges.
{\bf bottom right:} Distribution of the $Q$ factor after applying the air bridges. 
}
\label{fig5}
\end{figure}

We applied air-bridges in correspondence of each element of the array. Figure~\ref{fig5} (top) shows the details of the air-bridges that were applied to the same device characterized in Ref.~\cite{bullkid2022}. The left and right bottom panels shows the distribution of $Q$ before (Ref.~\cite{bullkid2022}) and after (this work) the modification. The 68\% dispersion improves from $1.3\times10^5$ to $0.7\times10^5$.

\section{SAMPLE PULSES ON THE NINE KIDs}\label{sec:appendix2}

The shape cuts $\chi^2_{L,R}$ in Eq.~\ref{eq:chi} are similar to those employed in other experiments, e.g. CUORE~\cite{CUORELowEnergyAnalysis} and NUCLEUS~\cite{CRAB2023}, and  efficiently identify pulses with shape different from the template if the signal to noise ratio (SNR) is sufficiently high. However, since pulse shape variables are evaluated including waveform samples with signal to noise ratio lower than the maximum, which is a proxy for the energy, their identification capability degrades rapidly while approaching the energy threshold. 

\begin{figure}[tb]
\includegraphics[width=1.0\linewidth]{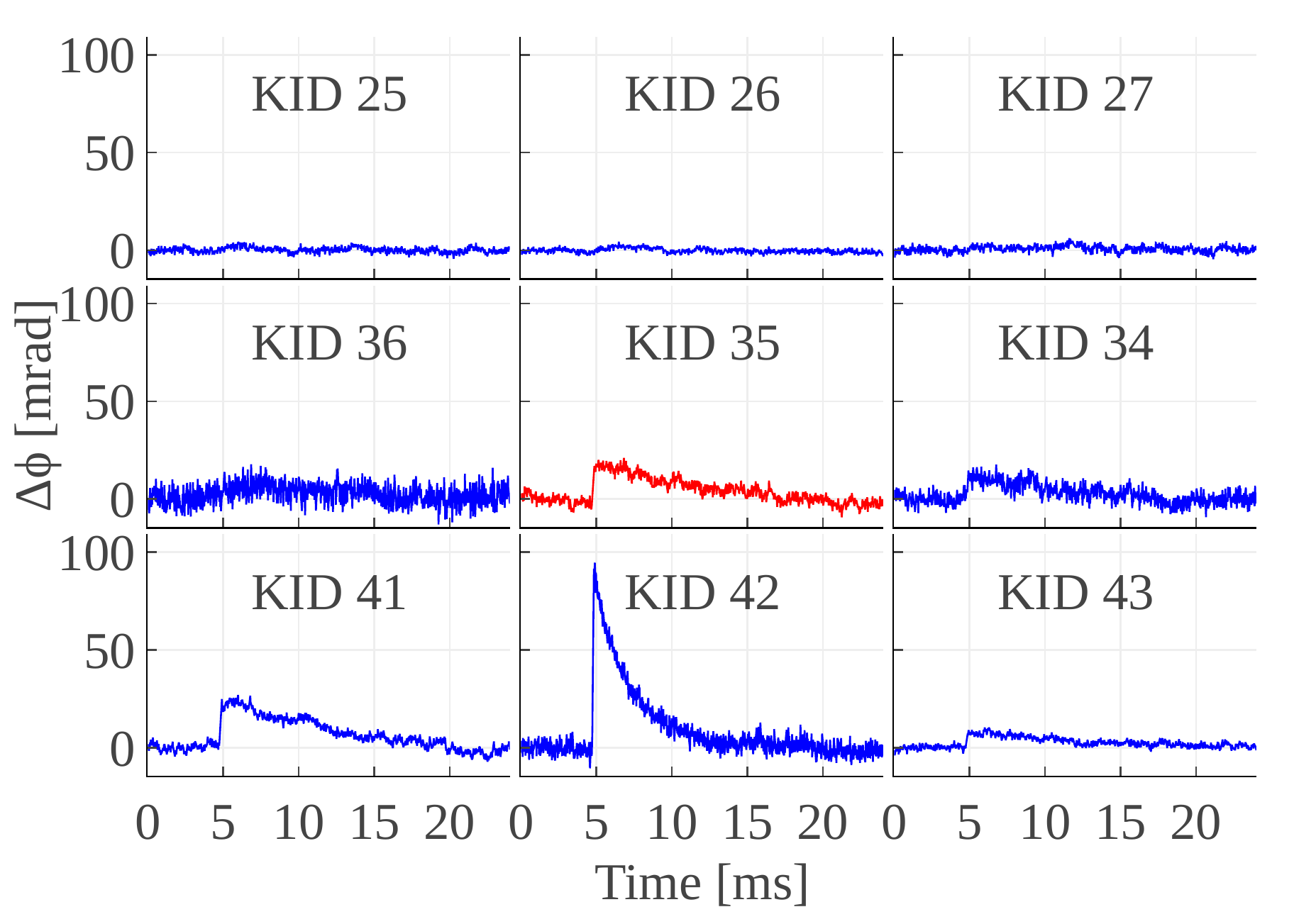}
\includegraphics[width=1.0\linewidth]{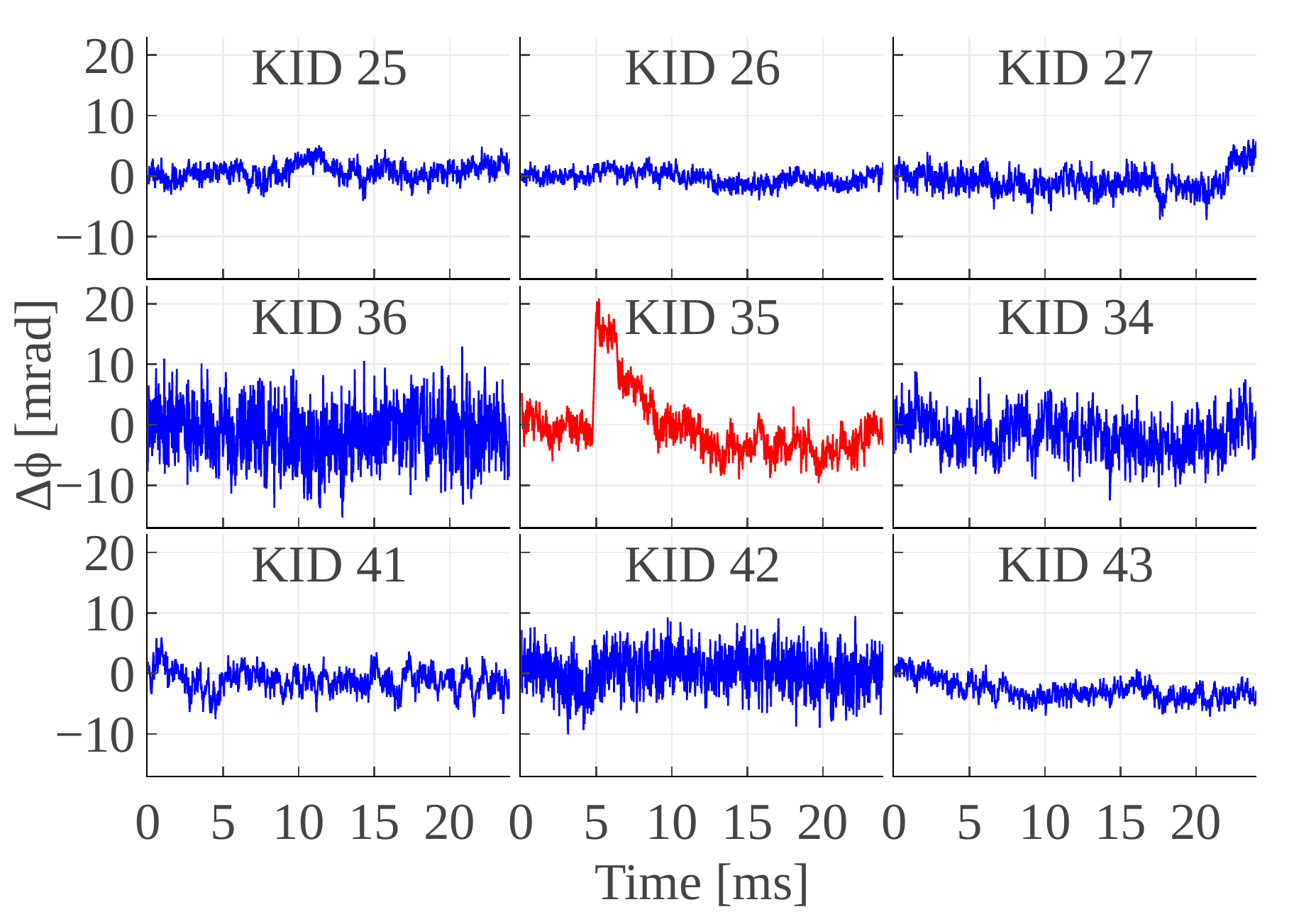}
\caption{{\bf Top:} Event passing the cuts on $\chi^2_{L,R}$ variables  and rejected by the cut on the $\psi_{1...8}$ variables; {\bf Bottom:} Event kept by the cuts on $\chi^2_{L,R}$ and $\psi_{1...8}$ variables.} 
\label{fig6}
\end{figure}

Figure~\ref{fig6} (top) shows a pulse of 500~eV which passes the $\chi^2_{L,R}$ cuts in the main KID (35), as its shape results compatible with that of a template at the given SNR. Looking at the side KIDs, however, it is clear that the event did not originate in KID 35, possibly in KID 42 or in a KID in even lower rows, while KID 35 is seeing the leaking phonons. When applying the cuts on the $\psi_{1...8}$ variables which include the amplitude of side KIDs (Eq.~\ref{eq:psi}), events like the one in the top panel are rejected while only events like the one in the bottom panel are retained.

\vspace{1cm}
\section{REVERSE TRIGGER SPECTRA}\label{sec:appendix3}

The reverse triggered data are processed following the same procedure of standard data. In the analysis a minus sign is applied to the (negatively triggered) data stream in order to use the same algorithms. The definition of the $\chi^2_{L,R}$ and $\psi_{1...8}$ is the same.

\begin{figure}[t]
\includegraphics[width=1.0\linewidth]{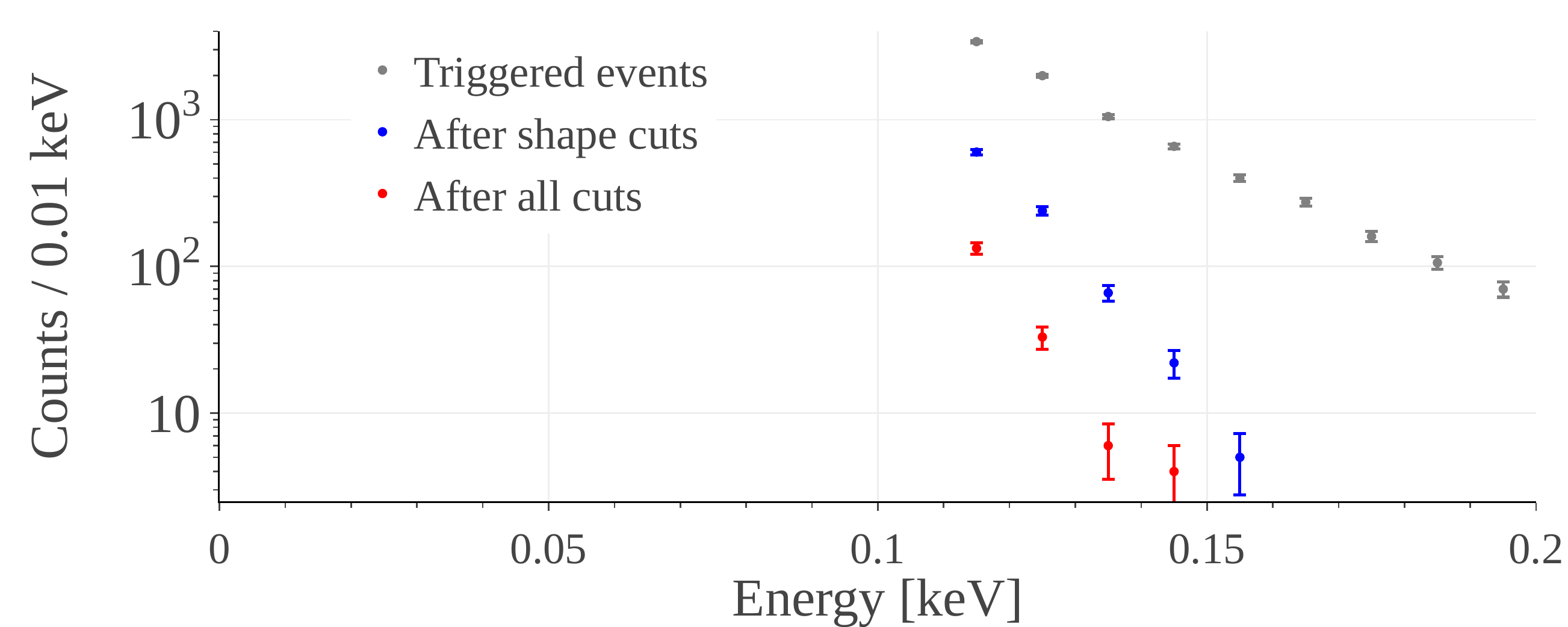}
\caption{Spectrum of reverse trigger events after cuts (same analysis of standard data).}
\label{fig7}
\end{figure}

Figure~\ref{fig7} shows the spectrum before cuts, after the application of the cuts on the $\chi^2_{L,R}$ variables and after the application of the cuts on the $\psi_{1...8}$ variables.
Given the small duration of the data taking, limited to only 30 minutes, the statistics after cuts is poor. The energy threshold for the analysis of standard data is set at 160~eV. It is however clear that the exponential tail of the  noise Gaussian can extend even to higher energies. More statistics of reverse triggered data will be acquired in the future, in order to better understand this source of background.

\section{COMPARISON WITH OTHER EXPERIMENTS}\label{sec:appendix4}
\begin{figure}[b]
\includegraphics[width=1\linewidth]{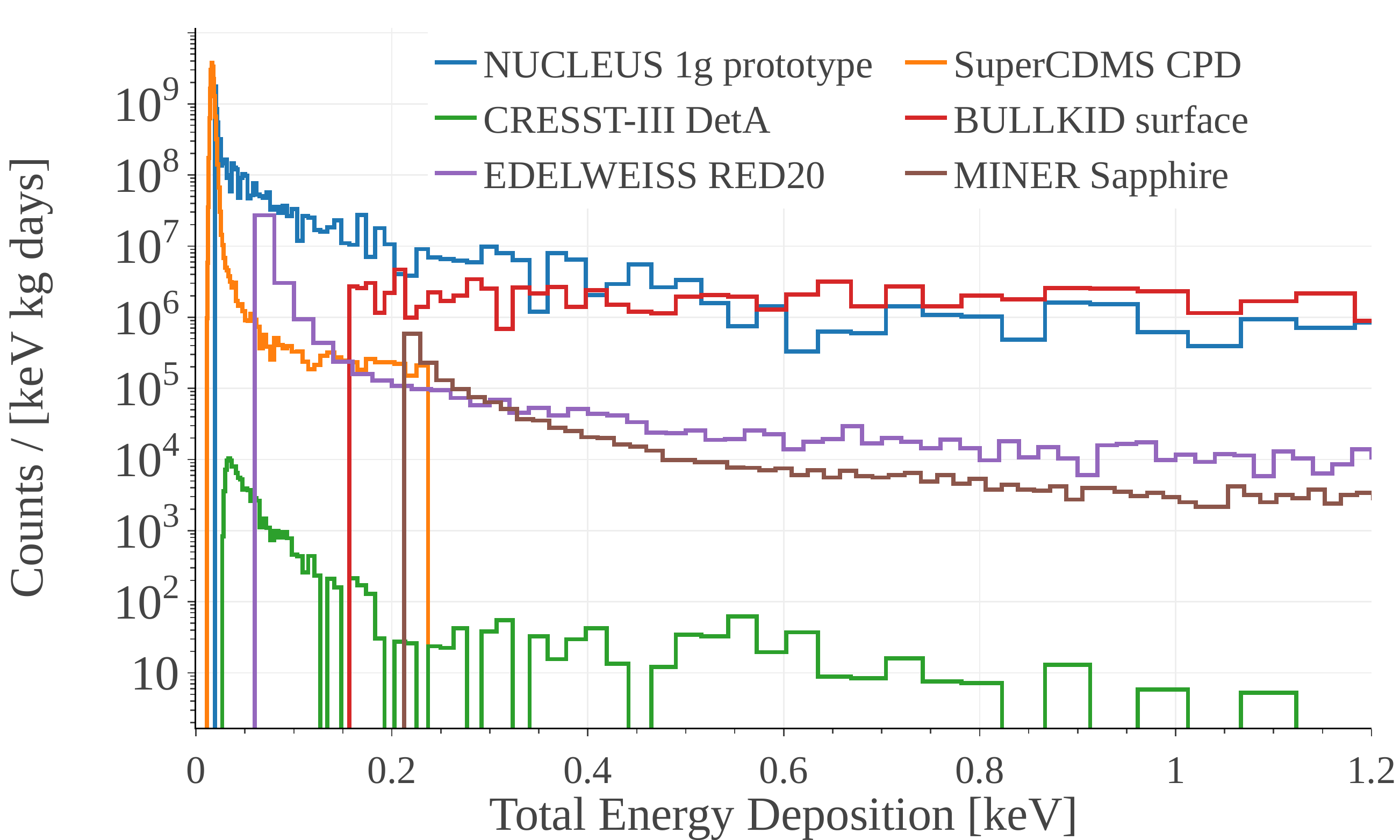}
\caption{Comparison of the BULLKID's spectrum presented in this work with other experiments of the field (see text). The data are taken from the public repository maintained by the organizers of the EXCESS workshop and plotted with their tool (See Ref. 15 in Ref.~\cite{excess2022}).}
\label{fig8}
\end{figure}

Figure~\ref{fig8} shows the spectrum presented in this work compared to other experiments in the field. It has to be stressed that experiments took data in different conditions: NUCLEUS 1~g was on surface like BULLKID but with an $^{55}$Fe source which might add background close to threshold; EDELWEISS RED20 and MINER used a lead shield which reduced the flat bacgkround by 2 orders of magnitude, and CRESST data were taken underground at Gran Sasso Laboratories in Italy with shielding.

\end{appendix}

\end{document}